\documentclass[a4paper,11pt]{iopart}
\pdfoutput=1 

\pdfoutput=1

\usepackage{graphicx}
\usepackage{bm}
\usepackage{braket}
\usepackage{amssymb}
\usepackage{cite}
\usepackage{mathrsfs}
\usepackage{bbm}
\usepackage{mathtools}
\usepackage{amsmath}
\usepackage{xcolor}
\usepackage{hyperref}
\usepackage{enumerate}
\usepackage[toc,title]{appendix}
\usepackage{verbatim}
\usepackage[margin=30pt, bf, font=footnotesize, center, justification=justified]{caption}[2004/07/16]
\usepackage[T1]{fontenc} 
\usepackage{xcolor}
\usepackage{subfigure}
\usepackage{braket}

\hypersetup{colorlinks,bookmarksopen,bookmarksnumbered,
citecolor=red,
linkcolor=blue,
pdfstartview=green,
urlcolor=teal}

\catcode`@=11 
\renewcommand\tableofcontents{%
  \section*{\contentsname}%
  \@starttoc{toc}%
}
\catcode`@=12

\begin{document}

\title[
]
{Generalized entanglement entropies in two-dimensional conformal field theory
}

\vspace{.5cm}

\author{Sara Murciano$^1$, Pasquale Calabrese$^{1,2}$ and Robert  M.  Konik$^3$}
\address{$^1$SISSA and INFN Sezione di Trieste, via Bonomea 265, 34136 Trieste, Italy.}
\address{$^{2}$International Centre for Theoretical Physics (ICTP), Strada Costiera 11, 34151 Trieste, Italy.}
\address{$^{3}$Condensed Matter Physics \& Materials Science Division,
 Brookhaven National Laboratory, Upton, NY 11973-5000, USA}

\vspace{.5cm}

\begin{abstract}
We introduce and study generalized R\'enyi entropies defined 
through the traces of products of  ${\rm Tr}_B (|\Psi_i\rangle\langle \Psi_j|)$ where  $|\Psi_i\rangle$ are eigenstates of a 
two-dimensional conformal field theory (CFT). 
When $|\Psi_i\rangle=|\Psi_j\rangle$ these objects reduce to 
the standard R\'enyi entropies of the eigenstates of the CFT.
Exploiting the path integral formalism, we show that the second  generalized R\'enyi entropies are equivalent to four point correlators.
We then focus on a free bosonic theory for which the mode expansion of the fields allows us to develop an efficient strategy to compute the second generalized R\'enyi entropy for all eigenstates. 
As a byproduct, our approach also leads to new results for the standard R\'enyi and relative entropies involving arbitrary descendent states of the bosonic CFT.

\end{abstract}

\maketitle

\newpage

\tableofcontents

\section{Introduction}
The characterization of the entanglement content of an extended quantum system is nowadays a widely studied research theme \cite{intro1, intro2, intro3}. A relevant contribution central to both high energy and condensed matter has come from two dimensional conformal field theory (CFT) that has led to a plethora of remarkable universal results for many entanglement related quantities. The most known ones are surely the R\'enyi entropies, which measure the entanglement between two complementary spatial regions, $A$ and $ B$, 
of an extended quantum system described by a pure state $|\Psi\rangle$ \cite{intro2,intro3}. They are defined as 
\begin{equation}
S_{A}^{(n)}= \frac{1}{1-n} \log \text{Tr} \rho_A^{n},
\end{equation}
where $\rho_A={\rm Tr}_{B} |\Psi\rangle\langle\Psi|$ is the reduced density matrix (RDM) associated to the subsystem $A$.
The von Neumann entropy is obtained as the limit for $n\to 1$ of $S_{A}^{(n)}$ and so
the R\'enyi entropies are a very useful theoretical tool, being the core of the replica approach to the entanglement  \cite{cc-04,cc-09}.
Furthermore R\'enyi entropies have been experimentally measured in cold atoms and ions trap setups \cite{exp1,exp2,lukin-18,bej-18,vek-21}.

An old and fundamental result is that the ground state entanglement entropy is universal  \cite{hlw-94,cc-04,cc-09}.
The same is also true for  all low-energy eigenstates \cite{sierra,afc-09}. 
Indeed, denoting by $\rho_{\Upsilon}$ the  RDM of the state associated to the field $\Upsilon$ by the operator-state correspondence (and hence $\rho_{\mathbbm{1}}$ is 
the ground-state RDM), the ratio 
\begin{equation} 
\label{eq:univ0}
F_{\Upsilon, n} (A) \equiv \frac{\text{Tr} \rho_{\Upsilon}^n}{ \text{Tr} \rho_{\mathbbm{1}}^n}=\exp[(1-n)(S^{(n)}_{A, \Upsilon}-S^{(n)}_{A, GS})] ,
\end{equation}
is universal and calculable in CFT \cite{sierra,berganza}. 
The universal function $F_{\Upsilon, n} (A)$ measures the excess of entanglement of the excited state  $\ket{\Upsilon}$ with respect to the ground state value.
The simplest class of excited states in a CFT are those generated by the action of a primary field $\Upsilon$. 
The von Neumann and R\'enyi entropies in excited states of CFT have then been the subject of intensive 
investigations \cite{sierra,berganza,jiaju1,jiaju2,jiaju3,jiaju4,jiaju5,cghw-15,palmai,palmai1,palmai2, Capizzi,elc-13,cel-14}.

Another important quantum information quantity encoding the universal features of CFT low-lying excited states is the relative entanglement entropy \cite{lashkari2016,lashkari2014, casini-2016, clt-16, bekenstein-bound, hol-rel-entropy,hol-rel-entropy2, ugajin2016, ugajin2016-2, ugajin-higherdim, abch-16,balasubramanian-14,mrc-19,paola,cc-21}.  
Given two RDMs, $\rho_1$ and $\rho_0$, the relative entropy is defined as  \cite{rel1, rel2}
\begin{equation} \label{relent}
S_{A}(\rho_1 \| \rho_0) = \text{Tr} \left( \rho_1 \log \rho_1 \right) -  \text{Tr} \left( \rho_1 \log \rho_0 \right),
\end{equation}
and can be interpreted as a measure of distinguishability of the two RDMs, providing a sort of distance between them in the Hilbert space (despite it not being a proper metric because it is not symmetric in $\rho_1$ and $\rho_0$).
In a replica approach, $S_{A}(\rho_1 \| \rho_0)$ can be obtained as the limit for $n\to1$ of the logarithm of the universal ratio \cite{lashkari2016,paola}
\begin{equation}
\label{eq:univrel0}
G_{n, A}(\rho_1||\rho_0) \equiv \dfrac{\mathrm{Tr}(\rho_1\rho_0^{n-1})}{\mathrm{Tr}\rho_1^n}.
\end{equation}
Similar quantities are also the starting point for the replica approach to the trace distance \cite{zrc-19,zrc-19b,zc-19}.
Both Eq. \eqref{eq:univ0} and Eq. \eqref{eq:univrel0} have been studied intensively for the lowest-energy states corresponding to primary fields. 
Conversely, the study of descendant states (obtained from the primaries by the application of conformal generators) has been limited to a very few cases \cite{palmai,palmai2}.
The drawback of the method of Refs. \cite{palmai,palmai2} is that it becomes more and more cumbersome as the conformal level increases.
Here we develop a strategy to obtain the R\'enyi entropies of excited states which is a more efficient way for theories with central charge $c=1$ for which we can use
the free boson representation. 

As part of this, we introduce a novel quantity that we dub  {\it generalized mixed state R\'enyi entropies} (GMSREs) defined as
\begin{equation}\label{eq:renyin}
R_{i_1,j_1;\dots; i_n,j_n}=\mathrm{Tr}_A \left(\prod_{t=1}^n \mathrm{Tr}_B \ket{\Psi_{i_t}}\bra{\Psi_{j_t}}\right),
\end{equation}
where $\ket{\Psi_{i_{t}}}$ is the $t$-th copy $(1 \leq t \leq n)$ of states $\ket{\psi_{i}}$ of the CFT.
For brevity, we will often refer to them as generalized R\'enyi entropies.
This object reduces to the usual R\'enyi entropies of excited states when $ \ket{\Psi_{i_t}}\bra{\Psi_{j_t}}=\delta_{i,j}  \ket{\Psi_{i}}\bra{\Psi_{j}}$ for each $t$-th copy. Let us stress that the states $\psi_{i_l}$ in each copy can be different CFT eigenstates, i.e. $\psi_{i_1}\neq \psi_{i_2}$. Moreover, the CFT eigenstates are also known in literature as dilatation eigenstates, since they are the eigenstates of the CFT Hamiltonian given by the Virasoro generators $L_0+\bar{L}_0$ (up to a constant), which is the dilatation operator.
These objects also resemble the notion of pseudo-entropies recently introduced in \cite{tt-05,tt-11}, whose reduced density matrix is defined in terms of two different states.
Apart from their intrinsic interest, these generalized non-diagonal entropies represent the building blocks for the calculation of the entanglement entropies in states
that can be written as a linear combination of the elements of the CFT basis, as happens for example, in the truncated conformal space approach \cite{tcsa,yz-91,mck-21,James_2018}. 

The paper is organised as follows. 
In Section \ref{sec2} we outline the path integral approach to the second generalized R\'enyi entropy and we show that it can be rewritten as a correlation function. 
In Section \ref{sec3} we focus on the bosonic theory and we develop a CFT approach to compute the second generalized entropy for an arbitrary state of the CFT. 
We start from some simple examples and piece by piece, we lead the reader to the most general case in Section \ref{sec:mgc}. 
Finally, we conclude and discuss some future perspectives in Section \ref{sec:concl}. Three appendices
are also included providing details about the analytical and numerical computations.

\section{Path integral for the second generalized R\'enyi entropy }\label{sec2}
\begin{figure}[t]
\centering
{\includegraphics[width=0.3\textwidth]{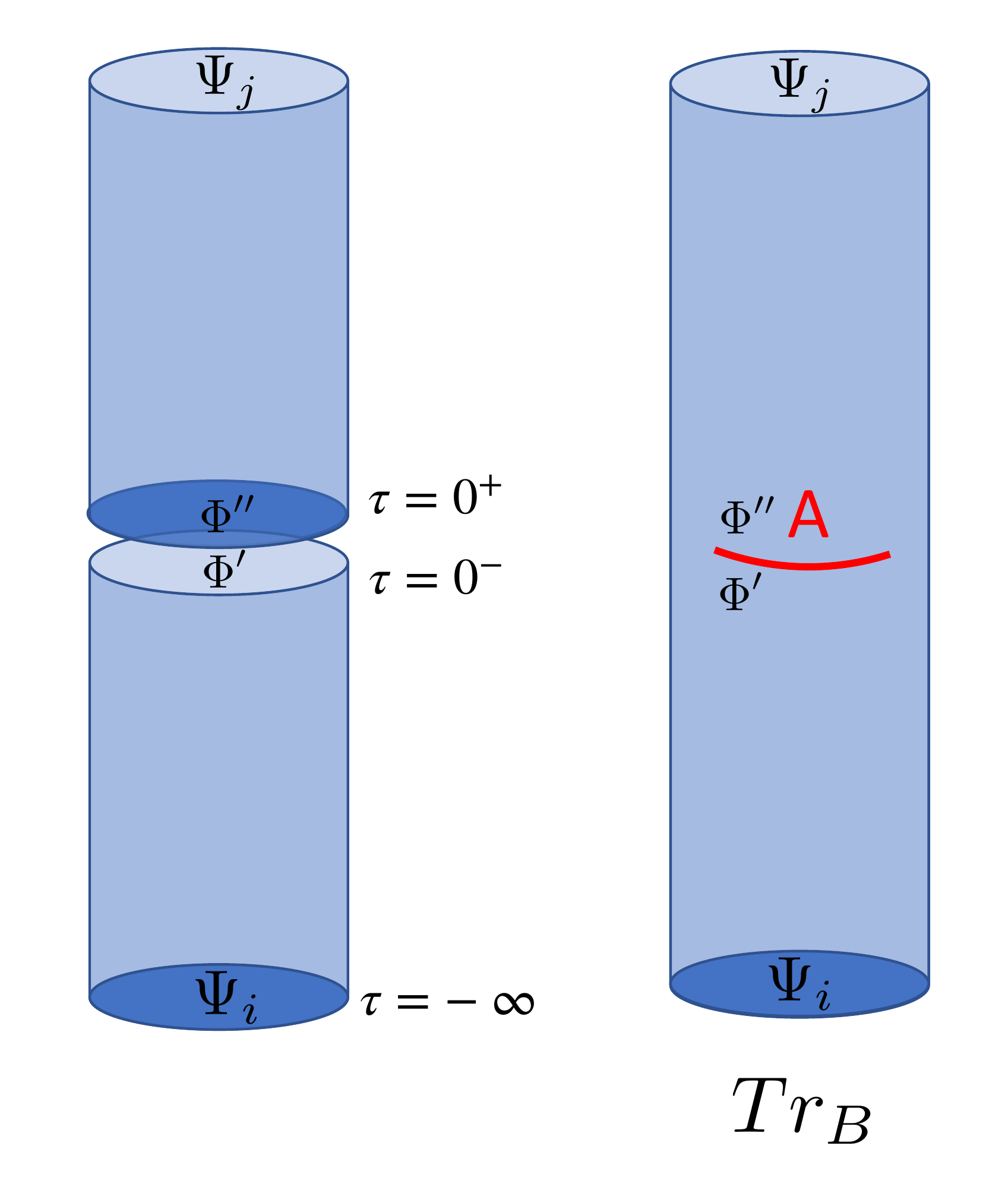}}
\caption{Left: Path integral representation of matrix elements of $\ket{\Psi_i}\bra{\Psi_j}$, cf. Eq. \eqref{eq:fig1}. Right: Path integral representation of $\mathrm{Tr}_B(\ket{\Psi_i}\bra{\Psi_j})$, as in Eq. \eqref{eq:off-rdm}.}
\label{fig:fig1}
\end{figure}

In this section, we focus on the second generalized R\'enyi entropy and we show that its calculation within the path integral formalism reduces to the computation of 
a four point correlator.  Let $R$ be the total length of a periodic 1D system, $A$ a subsystem consisting of a segment of length $\ell$, i.e $A = [u, v]$ with $v - u = \ell$, and $B$ its complement. 
Because of the state-operator correspondence in CFT, each eigenstate $|\Psi_{i_t}\rangle$ is generated by an operator $\Psi_{i_t}$, acting on the vacuum and placed 
at the infinite past.  
Hence, the path integral representation of the off-diagonal density matrix $\ket{\Psi_i}\bra{\Psi_j}$ is given, up to a normalization constant, by
 \begin{multline}\label{eq:fig1}
 \langle \Phi' \ket{\Psi_i}\langle\Psi_j\ket{\Phi''}\propto \int \mathcal{D}\phi e^{-S(\phi)}\prod_x(\delta(\phi(x,\tau=0^-)-\Phi'(x)))\Psi_i(\tau=-\infty) \\ \times \prod_x(\delta(\phi(x,\tau=0^+)-\Phi''(x)))\Psi_j(\tau=\infty),
 \end{multline}
 where the geometry of the integration surface is an infinite cylinder with a discontinuity of the field configurations at $\tau=0$ and $S(\phi)$ is the Euclidean action of the theory we are interested in (e.g. in the following we will focus on the bosonic CFT). We represent Eq. \eqref{eq:fig1} graphically in the left panel of Fig. \ref{fig:fig1}.

 Let us now compute the off-diagonal reduced density matrix by tracing over the subsystem $B$ (see the right panel of Fig. \ref{fig:fig1}):
 \begin{multline}\label{eq:off-rdm}
 \bra{\Phi'} \mathrm{Tr}_B(\ket{\Psi_i}\bra{\Psi_j} )\ket{\Phi''}= C_{ij} \int \mathcal{D}\phi e^{-S(\phi)}\prod_{x \in A}(\delta(\phi(x,\tau=0^-)-\Phi'(x)))\Psi_i(\tau=-\infty) \\ \times \prod_{x \in A}(\delta(\phi(x,\tau=0^+)-\Phi''(x)))\Psi_j(\tau=\infty),
 \end{multline} 
where $ C_{ij}$ is a normalization factor. By imposing the normalization of the off-diagonal reduced density matrix to be $\mathrm{Tr}_A(\mathrm{Tr}_B(\ket{\Psi_i}\bra{\Psi_j} ) )=\delta_{ij}$, we can fix the constant $C_{ij}$ as
\begin{equation}\label{eq:cij}
\delta_{ij}= C_{ij} Z \braket{\Psi_j (\tau = \infty)\Psi_i(\tau = -\infty)}.
\end{equation}
Here $Z$ and $\braket{\Psi_j (\tau = \infty)\Psi_i(\tau = -\infty)}$ are, respectively, the partition function and the correlation function of the two fields.

\begin{figure}[t]
\centering
{\includegraphics[width=0.4\textwidth]{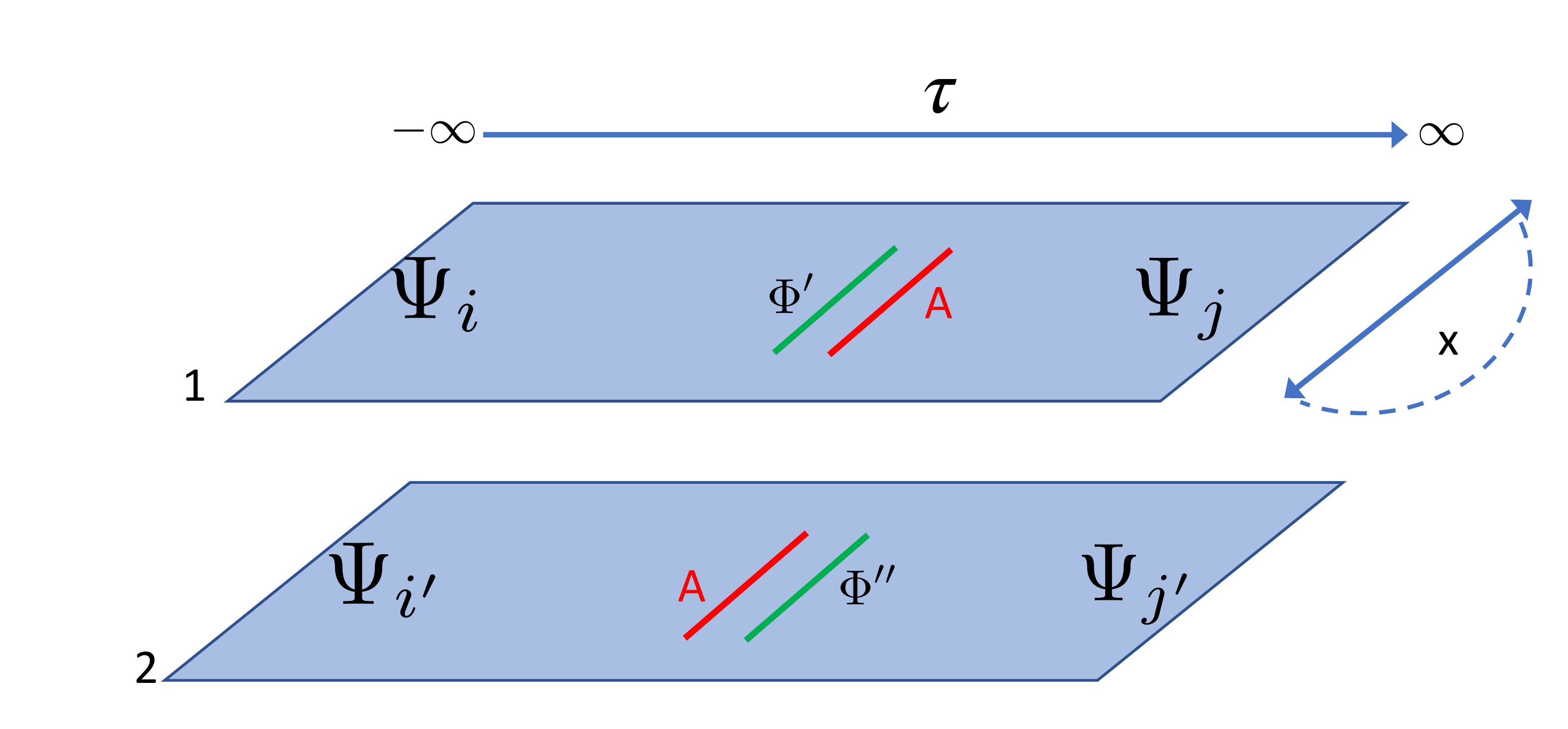}}
\caption{Representation of the spacetime geometry involving a 2-sheeted Riemann surface, $\mathcal{R}_2$, that
appears in computing the second generalized entropy $R_{i,j;i',j'}$ in Eq.~\eqref{eq:rii'}.}
\label{fig:fig2}
\end{figure}

As already mentioned, we are interested in the second generalized R\'enyi entropy. We thus make two copies of Eq. \eqref{eq:off-rdm}, and glue them together cyclically such that
\begin{equation}\label{eq:12}
\phi_1 (x,\tau=0^+) = \phi_2(x,\tau=0^-), \quad \phi_2 (x,\tau=0^+) = \phi_1(x,\tau=0^-),\quad \forall x \in A. 
\end{equation}
Starting from Eq. \eqref{eq:off-rdm}, we find the following expression for the elements of the off-diagonal second R\'enyi entropy
\begin{multline}\label{eq:off-rdm2}
 \bra{\Phi'} \mathrm{Tr}_B(\ket{\Psi_i}\bra{\Psi_j}) \mathrm{Tr}_B(\ket{\Psi_{i'}}\bra{\Psi_{j'}} )\ket{\Phi''}= C_{ij}C_{i'j'} \int \mathcal{D}\phi_1 \mathcal{D}\phi_2 e^{-S(\phi_1)-S(\phi_2)}\\ \times \prod_{x \in A}(\delta(\phi_1(x,\tau=0^-)-\Phi'(x)))   \prod_{x \in A}(\delta(\phi_2(x,\tau=0^+)-\Phi''(x)))\\ \times \Psi_i(\tau=-\infty)\Psi_j(\tau=\infty)\Psi_{i'}(\tau=-\infty)\Psi_{j'}(\tau=\infty),
 \end{multline} 
 where the path-integral is done over a 2-sheeted Riemann surface such that the first equality in Eq. \eqref{eq:12} holds. Let us notice that the field $\phi_1$ along $A$ at $\tau=0^-$ is fixed at $\Phi'$, while $\phi_2$ at $\tau=0^+$ is fixed at $\Phi''$.
 Thus, by taking the trace over $A$, i.e. using also the second identification in Eq. \eqref{eq:12}, we obtain an expression for the second generalized R\'enyi entropy
 \begin{multline}\label{eq:rii'}
 R_{i,j;i',j'}=\mathrm{Tr}_A(\mathrm{Tr}_B(\ket{\Psi_i}\bra{\Psi_j}) \mathrm{Tr}_B(\ket{\Psi_{i'}}\bra{\Psi_{j'}} ))= C_{ij}C_{i'j'} \int \mathcal{D}\phi_1 \mathcal{D}\phi_2 e^{-S(\phi_1)-S(\phi_2)} \\ \times \prod_{x \in A}\delta(\phi_1(x,\tau=0^-)-\Phi'(x)) \delta(\phi_2(x,\tau=0^+)-\Phi''(x))\Psi_i(\tau=-\infty)\Psi_j(\tau=\infty)  \Psi_{i'}(\tau=-\infty) \\ \times \Psi_{j'}(t=\infty)= C_{ij}C_{i'j'}Z_2(A)\braket{\Psi_i(\tau=-\infty)\Psi_j(\tau=\infty)\Psi_{i'}(\tau=-\infty)\Psi_{j'}(\tau=\infty)}_{\mathcal{R}_2},
 \end{multline}
 where $\mathcal{R}_2$ is the 2-sheeted Riemann surface that results from the sewing of the two copies of the original cylinder along the cuts associated with the interval $A$ and $Z_2(A)$ is the partition function in the same geometry. 
This path integral is over a spacetime pictured in Fig. \ref{fig:fig2}. The result in Eq. \eqref{eq:rii'} shows that the second generalized entropy reduces to a 4-point correlation function.  

Before embarking in the calculation of Eq. \eqref{eq:rii'}, we fix the normalization constants, $C_{ij}$'s. 
As already discussed, the constants $C_{ij}$'s are independent of $A$.
Thus if the segment $A$ shrinks to zero, we decouple the two Riemann sheets and the 4-point function factorizes into a product of 2-point ones:
 \begin{equation}\label{eq:ato0}
 C_{ij}C_{i'j'}Z^2\braket{\Psi_i(\tau=-\infty)\Psi_j(\tau=\infty)}_{\mathcal{R}_1}\braket{\Psi_{i'}(\tau=-\infty)\Psi_{j'}(\tau=\infty)}_{\mathcal{R}_1}=\delta_{ij}\delta_{i'j'}.
 \end{equation}
 Now if we take the segment $A$ to be the entire system (as the $\mathrm{Tr}_B$ operation is trivial), the two sheets of the Riemann surface are again no longer connected and 
  \begin{equation}
 C_{ij}C_{i'j'}Z^2\braket{\Psi_i(\tau=-\infty)\Psi_{j'}(\tau=\infty)}_{\mathcal{R}_1}\braket{\Psi_{i'}(\tau=-\infty)\Psi_{j}(\tau=\infty)}_{\mathcal{R}_1}=\delta_{ij'}\delta_{i'j},
 \end{equation}
 i.e. the four-point function still factorizes even though differently with respect to Eq. \eqref{eq:ato0}. Hence, the constant may be fixed as 
 \begin{equation}
  C_{ij}=\frac{1}{Z \sqrt{\braket{\Psi_i(\tau=-\infty)\Psi_{i}(\tau=\infty)}_{\mathcal{R}_1}\braket{\Psi_{j}(\tau=-\infty)\Psi_{j}(\tau=\infty)}_{\mathcal{R}_1}}},
 \end{equation}
 and we can express ratios of the generalized entropies in terms of universal quantities:
 \begin{equation}\label{eq:Ffactors}
 \begin{split}
 \frac{R_{i,j;i',j'}}{R_{\mathbbm{1},\mathbbm{1};\mathbbm{1},\mathbbm{1}}}=&F_iF_jF_{i'}F_{j'} \braket{\Psi_i(\tau=-\infty)\Psi_j(\tau=\infty)\Psi_{i'}(\tau=-\infty)\Psi_{j'}(\tau=\infty)}_{\mathcal{R}_2},\\
 F_i=&\frac{1}{\sqrt{\braket{\Psi_i(\tau=-\infty)\Psi_{i}(\tau=\infty)}_{\mathcal{R}_1}}}.
 \end{split}
 \end{equation}
 This expression can be easily extended to the $n$-th order generalized R\'enyi entropy introduced in Eq. \eqref{eq:renyin} as
 \begin{multline}
\frac{R_{i_1,j_1;\dots; i_n,j_n}}{{R_{\mathbbm{1},\mathbbm{1};\dots; \mathbbm{1},\mathbbm{1}}}}\\=F_{i_1}F_{j_1}\dots F_{i_n}F_{j_n} \braket{\Psi_{i_1}(\tau=-\infty)\Psi_{j_1}(\tau=\infty)\dots\Psi_{i_n}(\tau=-\infty)\Psi_{j_n}(\tau=\infty)}_{\mathcal{R}_n}.
\end{multline}
  \section{Bosonic theory: a CFT approach beyond primary fields}\label{sec3}
 In the following sections, we compute the second generalized R\'enyi entropy for a  massless free bosonic field theory whose Euclidean action is
\begin{equation}\label{eq:action}
S(\phi)=\int_0^R dx d\tau \left[\frac{1}{8\pi}\partial_{\mu}\phi(x,\tau)\partial^{\mu}\phi(x,\tau) \right].
\end{equation}
The mode expansion of the field $\phi(x,\tau)$ is \cite{difrancesco}
 \begin{equation}\label{eq:Fexp}
 \phi(x,\tau)=\phi_0-i\frac{4\pi}{R}\pi_0\tau+i\sum_{k\neq 0}\frac{1}{k}(a_ke^{\frac{2\pi k(ix-\tau)}{R}}-\bar{a}_{-k}e^{\frac{2\pi  k (ix+\tau)}{R}}).
 \end{equation}
Here in Eq. \eqref{eq:Fexp} $\phi_0$ is the bosonic zero mode and $\pi_0$ is its conjugate momenta.
The modes $a_{-k},\bar a_{-k}$ can be obtained as follows:
 \begin{align}
a_{-k} &= -\frac{e^{-\frac{2\pi k\tau}{R}}}{2\pi} \int dxe^{\frac{2\pi i k }{R}x}\partial_{\omega}\phi,\quad  \omega=x+i\tau;\\
\bar{a}_{-k} &= \frac{e^{-\frac{2\pi k\tau}{R}}}{2\pi} \int dxe^{-\frac{2\pi i k }{R}x}\partial_{\bar{\omega}}\phi,\quad \bar{\omega}=x-i\tau.
 \end{align}
Each mode $a_{-k},\bar a_{-k}$ carries momentum $\pm 2\pi k/R$.

In Refs. \cite{sierra,paola},  Eq. \eqref{eq:univ0} and Eq. \eqref{eq:univrel0}, respectively, have been computed for the excited states generated by primary fields of a free massless bosonic field described by Eq. \eqref{eq:action}. The primary fields of the theory consists just of the vertex operator, $V_{\alpha,\bar{\alpha}}$, and the derivative field $i\partial_{\omega} \phi$ \cite{difrancesco}.  
%
The general state can be written in terms of the modes as
\begin{equation}
\ket{\Psi_i}=a_{-k_1}\dots a_{-k_i}\ket{M_i}, \quad \ket{M_i}=:e^{iM_i\beta \phi(0)}:\ket{0},
\end{equation}
with $\ket{M_i}$ being the highest weight states and $:e^{iM_i\beta \phi(0)}:$ is a vertex operator ($::$ denotes the normal ordering prescription in its mode expansion, 
i.e. positive modes should appear to the right of negative ones).  Here $\beta$ marks the compactification radius, $\beta^{-1}$, of the boson:
\begin{equation}
    \phi(x+R) \equiv \phi(x) + 2\pi\beta^{-1},
\end{equation}
and $M_i$ is an integer.
To treat a non-compact boson, one would likely have to choose a zeromode basis other than plane waves.
 In the following sections, we develop an efficient strategy to compute the generalized R\'enyi entropies involving  descendant states.

 \subsection{A warmup: zero-momentum states with a pair of chiral and anti-chiral modes}

As a first non-trivial example, let us evaluate the second generalized R\'enyi entropy for the following (normalized) states that involve a chiral and anti-chiral mode whose momentum sums to zero:
\begin{equation}\label{eq:warmup}
R_{k_1,k_2;k_3,k_4}=\mathrm{Tr}_A(\mathrm{Tr}_B(\frac{1}{k_1}a_{-k_1}\bar{a}_{-k_1}\ket{0}\bra{0}\frac{1}{k_2} a_{k_2}\bar{a}_{k_2})\mathrm{Tr}_B(\frac{1}{k_3}a_{-k_3}\bar{a}_{-k_3}\ket{0}\bra{0}\frac{1}{k_4} a_{k_4}\bar{a}_{k_4})).
\end{equation}
We can write the state $\frac{1}{k}a_{-k}\bar{a}_{-k}\ket{0}$ as
\begin{equation}
\frac{1}{k}a_{-k}\bar{a}_{-k}\ket{0}=-\frac{1}{4\pi^2 k}e^{-\frac{4\pi k \tau_{-}}{R}}\int_0^Rdx_1dx_{1'}e^{\frac{2\pi i k}{R}(x_1-x_{1'})}\partial_{\omega_1}\phi(\omega_1)\partial_{\bar{\omega}_{1'}}\phi(\bar{\omega}_{1'})\ket{0},
\end{equation}
where $\omega_{1,1'}=x_{1,1'}+i\tau_{-}, \tau_{\pm}=\pm \infty$. Since these states are normalized, the $F$-factors of Eq. \eqref{eq:Ffactors} are simply $1$. With this representation, we can write $R_{k_1,k_2;k_3,k_4}$ as 
\begin{equation}\label{eq:I's}
\begin{split}
\frac{R_{k_1,k_2;k_3,k_4}}{R_{\mathbbm{1},\mathbbm{1};\mathbbm{1},\mathbbm{1}}}&=\frac{1}{256 \pi^8 \prod_{i=1}^4k_i}e^{\frac{4\pi}{R}(k_2+k_4)\tau_+}e^{-\frac{4\pi}{R}(k_1+k_3)\tau_-}I_{1234}\bar{I}_{1'2'3'4'},\\
I_{1234}&=\displaystyle \int_0^R dx_1dx_2dx_3dx_4 e^{\frac{2\pi i }{R}(k_1x_1+k_3x_3-k_2x_2-k_4x_4)}\braket{\partial \phi (\omega_1)\partial\phi(\omega_2)\partial\phi(\omega_3)\partial\phi(\omega_4)}, \\
\bar{I}_{1'2'3'4'}&=\displaystyle \int_0^R dx_{1'}dx_{2'}dx_{3'}dx_{4'} e^{-\frac{2\pi i }{R}(k_1x_{1'}+k_3x_{3'}-k_2x_{2'}-k_4x_{4'})}\braket{\partial \bar{\phi} (\bar{\omega}_{1'})\partial\bar{\phi}(\bar{\omega}_{2'})\partial\bar{\phi}(\bar{\omega}_{3'})\partial\bar{\phi}(\bar{\omega}_{4'})}\\ &= I_{1234}^*,
\end{split}
\end{equation}
where $w_{1,1'}=x_{1,1'}+i\tau_-$ and $w_{2,2'}=x_{2,2'}+i\tau_+$ are coordinates on the first Riemann sheet while $w_{3,3'}=x_{3,3'}+i\tau_-$ and $w_{4,4'}=x_{4,4'}+i\tau_+$ are coordinates on the second Riemann sheet. 

Let us next compute the correlators appearing in Eq. \eqref{eq:I's}.  The complex coordinate $\omega=x+i\tau$ parametrizes an infinite cylinder of length $R$ and the interval $A$ is identified with the domain $u < \omega < v$. This cylinder can be mapped into the complex plane by the conformal transformation
\begin{equation}
\xi=\frac{\sin(\pi(\omega - u)/R)}{\sin(\pi(\omega -v)/R)},
\end{equation}
which maps the segment $( u,v)$ into  $(-\infty,0)$. Then, the uniformizing map for $\mathcal{R}_2$, $z=\xi^{1/2}$ tranforms the double-sheeted surface into a single sheet. 
As a consequence, the coordinates become
\begin{align}
\omega_{1,3} =&x_{1,3}+  i\tau_-\to  \xi_{-} = e^{i\pi(v-u)/R} \equiv e^{i\pi r} \to  z_{1} = e^{i\pi r/2}, z_{3} =- e^{i\pi r/2},\\
\omega_{2,4} =& x_{2,4}+ i\tau_+\to  \xi_{+} = e^{-i\pi(v-u)/R} \equiv e^{-i\pi r} \to  z_{2} = e^{-i\pi r/2}, z_{4} =- e^{-i\pi r/2},
\end{align}
each of the points $\xi_{-}$ and $\xi_{+}$ gives rise to two points in the Riemann surface $\mathcal{R}_2$ with coordinates $z_1,z_3$ and $z_2,z_4$, respectively. 
Thus, we can write $I_{1234}$ as ($z_{ij}\equiv z_i-z_j$)
\begin{equation}\label{eq:4d}
\begin{split}
 I_{1234}&=\displaystyle \int_0^R dx_1dx_2dx_3dx_4 e^{\frac{2\pi i }{R}(k_1x_1+k_3x_3-k_2x_2-k_4x_4)} \frac{dz}{d\omega}\Big|_{\omega=\omega_1}\frac{dz}{d\omega}\Big|_{\omega=\omega_2}\frac{dz}{d\omega}\Big|_{\omega=\omega_3}\frac{dz}{d\omega}\Big|_{\omega=\omega_4} \\
&\left( \frac{1}{z_{23}^2z_{14}^2}+\frac{1}{z_{12}^2z_{34}^2}+\frac{1}{z_{13}^2z_{24}^2} \right)=\\
&=\displaystyle \oint_{C_-^1}\oint_{C_+^2} \oint_{C_-^3}\oint_{C_+^4}dz_1dz_2dz_3dz_4e^{\frac{2\pi i }{R}(k_1x_1+k_3x_3-k_2x_2-k_4x_4)} \left( \frac{1}{z_{23}^2z_{14}^2}+\frac{1}{z_{12}^2z_{34}^2}+\frac{1}{z_{13}^2z_{24}^2} \right),
\end{split}
\end{equation}
where $C_-^{1,3}$ is a clockwise contour about $z_{1,3}=\pm e^{\pi i r/2}$ and $C_+^{2,4}$ is a counter clockwise contour about  $z_{2,4}=\pm e^{-\pi i r/2}$.
Using
\begin{equation}
\begin{split}\label{eq:points2}
e^{2\pi i x_{1,3}/R} &=e^{2\pi \tau_-/R}e^{2\pi i v/R}\frac{z_{1,3}^2-e^{-i\pi r}}{z_{1,3}^2-e^{i\pi r}},\\
e^{2\pi i x_{2,4}/R} &=e^{-2\pi \tau_+/R}e^{-2\pi i v/R}\frac{z_{2,4}^2-e^{i\pi r}}{z_{2,4}^2-e^{-i\pi r}},
\end{split}
\end{equation}
we can evaluate 
\begin{equation}
\begin{split}
& I_{1234}=e^{\frac{2\pi}{R}((k_1+k_3)\tau_- -(k_2+k_4)\tau_+)} \frac{16\pi^4 }{\prod_{i=1}^4\Gamma(k_i)}S(r,k_1,k_2,k_3,k_4),\\
& S(r,k_1,k_2,k_3,k_4) \equiv \partial_{z_1}^{k_1-1}\partial_{z_2}^{k_2-1}\partial_{z_3}^{k_3-1}\partial_{z_4}^{k_4-1}\left[\left(\frac{z_{1}^2-e^{-i\pi r}}{z_{1}+e^{i\pi r/2}} \right)^{k_1}\left(\frac{z_{2}^2-e^{i\pi r}}{z_{2}+e^{-i\pi r/2}} \right)^{k_2} \right. \\
&\left(\frac{z_{3}^2-e^{-i\pi r}}{z_{3}-e^{i\pi r/2}} \right)^{k_3}\left(\frac{z_{4}^2-e^{i\pi r}}{z_{4}-e^{-i\pi r/2}} \right)^{k_4} 
\left. \left( \frac{1}{z_{23}^2z_{14}^2}+\frac{1}{z_{12}^2z_{34}^2}+\frac{1}{z_{13}^2z_{24}^2} \right)\right]
\Bigg|_{\substack{z_{1,3}=\pm e^{i\pi r/2}  \\ z_{2,4}=\pm e^{-i\pi r/2}}}, 
\end{split}
\end{equation}
where $\Gamma$ denotes the Gamma function.
Therefore the second generalized entropy in Eq. \eqref{eq:warmup} reads
\begin{equation}\label{eq:Rk}
\frac{R_{k_1,k_2;k_3,k_4}}{R_{\mathbbm{1},\mathbbm{1},\mathbbm{1},\mathbbm{1}}}=\frac{1}{\prod_{i=1}^4k_i \Gamma(k_i)^2}|S(r,k_1,k_2,k_3,k_4)|^2.
\end{equation}
For $k_i = 1, i=1,\dots,4$, it reduces to
\begin{equation}\label{eq:sierra}
\frac{R_{1,1;1,1}}{R_{\mathbbm{1},\mathbbm{1};\mathbbm{1},\mathbbm{1}}}=\frac{(7+\cos(2\pi r))^4}{64^2}.
\end{equation}
This result reproduces what has been found in \cite{sierra} for the (chiral) primary field $i\partial \phi$. 
 In a similar way, we can compute 
\begin{equation}
\begin{split}\label{eq:results}
&\frac{R_{0,1;1,0}}{R_{\mathbbm{1},\mathbbm{1};\mathbbm{1},\mathbbm{1}}}=\frac{R_{1,0;0,1}}{R_{\mathbbm{1},\mathbbm{1};\mathbbm{1},\mathbbm{1}}}=\frac{\sin^4 \pi r }{16 \cos^4( \pi r/2)}, \\
&\frac{R_{1,1;0,0}}{R_{\mathbbm{1},\mathbbm{1};\mathbbm{1},\mathbbm{1}}}=\frac{R_{0,0;1,1}}{R_{\mathbbm{1},\mathbbm{1};\mathbbm{1},\mathbbm{1}}}=\frac{\sin^4 \pi r }{16 \sin^4( \pi r/2)},\\
&\frac{R_{1,0;1,0}}{R_{\mathbbm{1},\mathbbm{1};\mathbbm{1},\mathbbm{1}}}=\frac{\sin^4 (\pi r)}{16 }.
\end{split}
\end{equation}
The quantity $R_{0,0;1,1}/R_{\mathbbm{1},\mathbbm{1};\mathbbm{1},\mathbbm{1}}$ in Eq.\,(\ref{eq:results}) reproduces the results found for the relative entropy of the state described by the primary operator $i\partial \phi$ with respect to the ground state, $G(\rho_{GS}||\rho_{i\partial \phi})$, in Ref. \cite{paola}. 

\begin{figure}
\centering
\subfigure
{\includegraphics[width=0.48\textwidth]{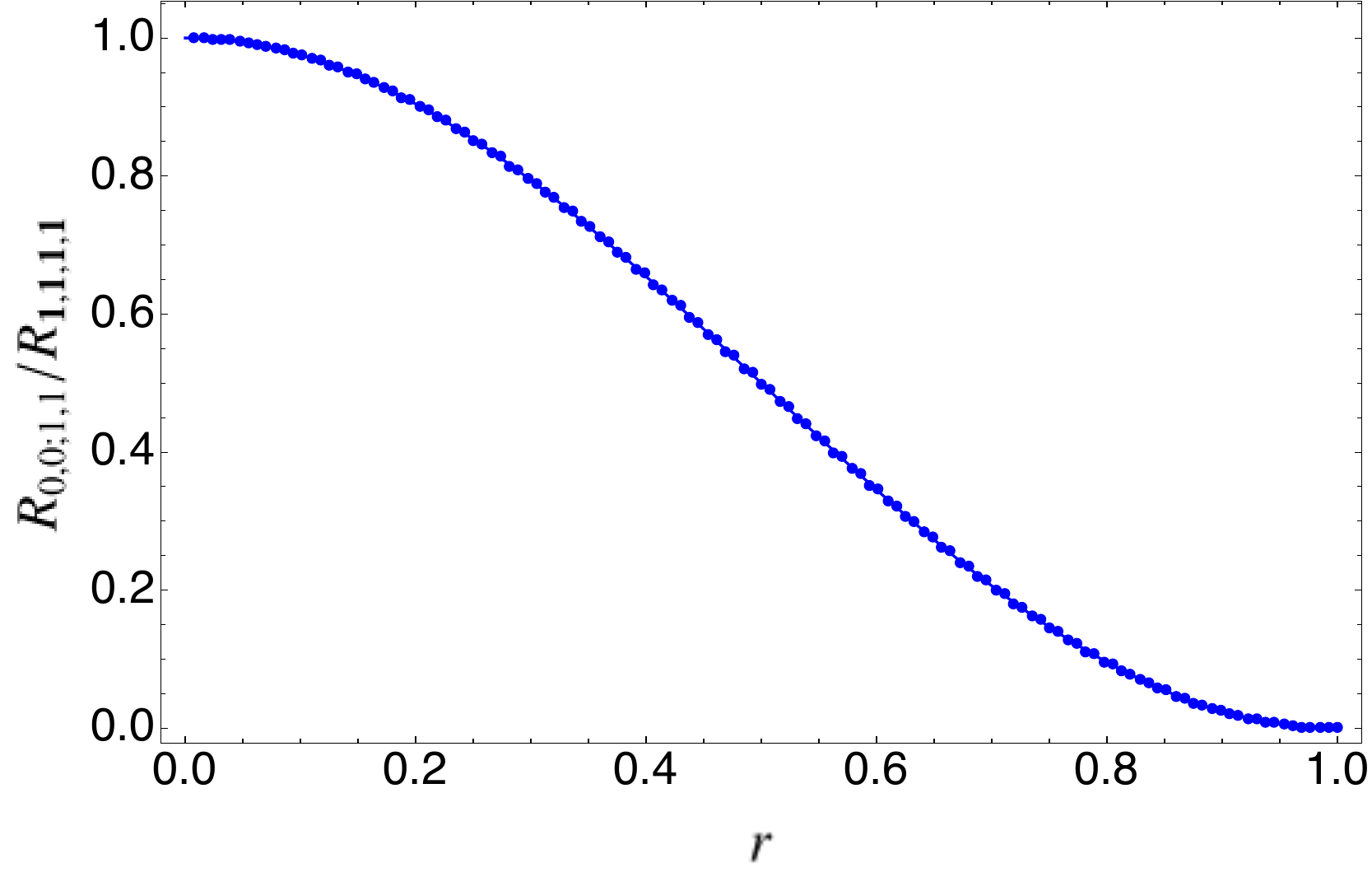}}
\subfigure
   {\includegraphics[width=0.48\textwidth]{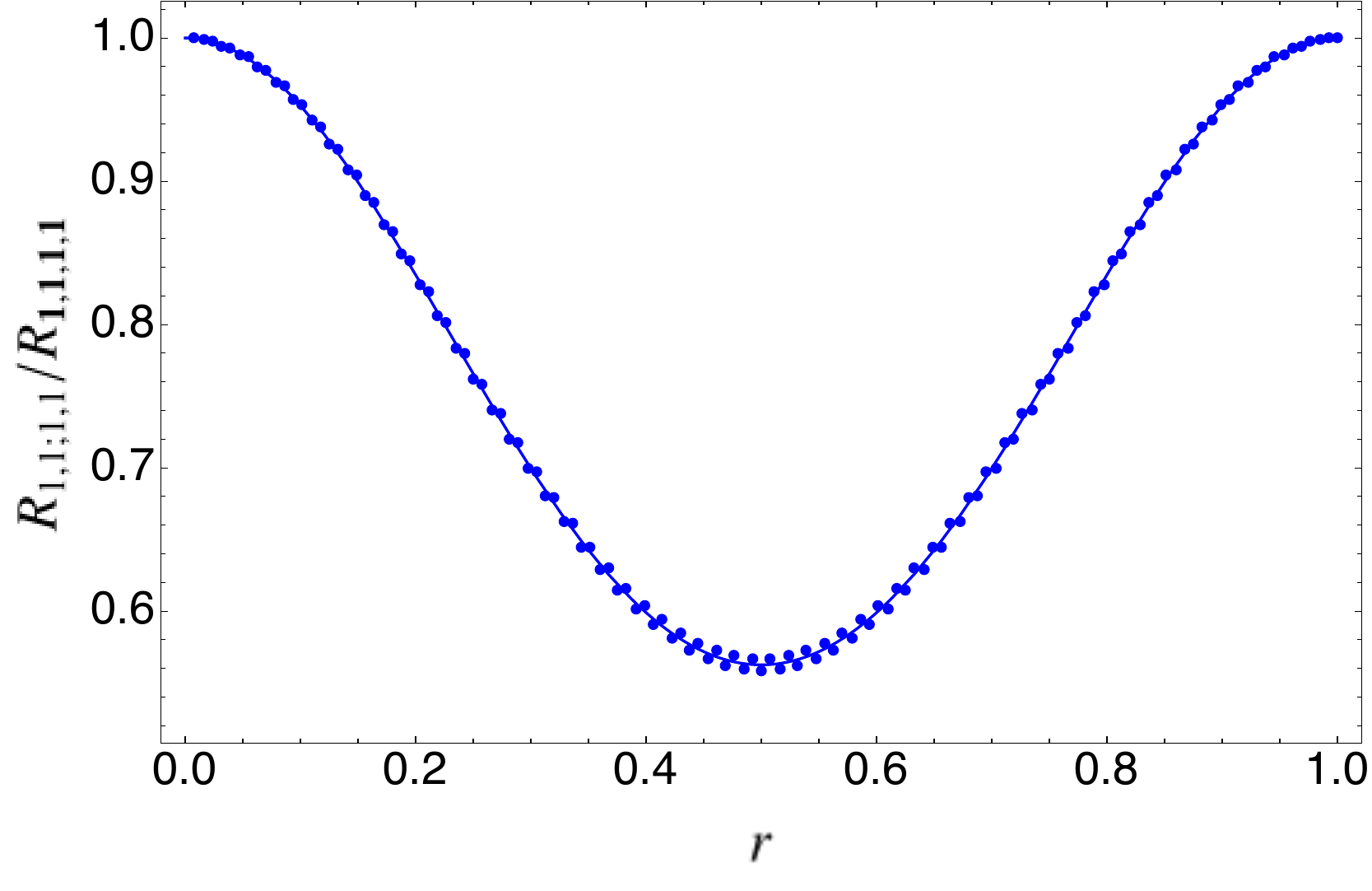}}
\caption{Tests of the CFT results for the generalized R\'enyi entropies in the XX spin chain. We consider spin chains made of $N=128$ spins. 
Left panel: The quantity $\frac{R_{0,0;1,1}}{R_{\mathbbm{1},\mathbbm{1},\mathbbm{1},\mathbbm{1}}}$ as a function of $r=\ell/N$. 
The symbols are the numerical data, while the continuous line is the square root of Eq. \eqref{eq:results}. 
Right panel:  $\frac{R_{1,1;1,1}}{R_{\mathbbm{1},\mathbbm{1},\mathbbm{1},\mathbbm{1}}}$ as a function of $r=\ell/N$. 
The analytical prediction is the square root of Eq.~\eqref{eq:sierra}.}
\label{fig:p-gs}
\end{figure}

We test this prediction and Eq. \eqref{eq:sierra} against exact numerical calculations in the XX spin-chain, as showed in Fig. \ref{fig:p-gs}, for a state involving the chiral component of the bosonic field. The XX chain can be mapped onto a lattice free-fermionic theory and then, in the continuum limit, to a massless Dirac fermion whose low-energy regime can be formulated, via a bosonization, as the free bosonic theory in Eq. \eqref{eq:action}. The explicit correspondence between microscopic low energy excitations of the XX chain and the primary operators of the compact boson has been discussed in Ref. \cite{sierra}. In particular, the primary operator $i\partial \phi$ is associated to the particle-hole excitation \cite{sierra} as is briefly reviewed in Appendix \ref{appendixTools}. Since the considered excitations are chiral, we have to take the square roots of the previous expressions. 

\subsection{An example involving vertex operators}

As a second non-trivial example,  we choose states involving both the derivative operator $i\partial \phi$ and vertex operators.  In particular,
we consider one of the states to be 
\begin{equation}
 V_{\alpha,\bar{\alpha}}(\omega,\bar{\omega})\ket{0}= (\frac{iR}{2\pi}e^{2\pi i/R\omega})^{\alpha^2/2} (\frac{iR}{2\pi}e^{2\pi i/R\bar{\omega}})^{\bar{\alpha}^2/2}:e^{i(\alpha \phi (\omega)+\bar{\alpha}\bar{\phi}(\bar{\omega}))}:\ket{0},\quad  \omega=i\tau_-.
\end{equation}
Therefore the second generalized R\'enyi entropy we compute is
\begin{equation}\label{eq:eg}
R_{k_1,k_2;V_{\alpha,\bar{\alpha}},V_{-\alpha,-\bar{\alpha}}}=\mathrm{Tr}_A(\mathrm{Tr}_B(\frac{1}{k_1}a_{-k_1}\bar{a}_{-k_1}\ket{0}\bra{0}\frac{1}{k_2} a_{k_2}\bar{a}_{k_2})\mathrm{Tr}_B(V_{\alpha,\bar{\alpha}}\ket{0}\bra{0}V_{-\alpha,-\bar{\alpha}})).
\end{equation}
If we rename $V_{\alpha}$ and $V_{\bar{\alpha}}$ the holomorphic and antiholomorphic part of the vertex operator, respectively, Eq.\,(\ref{eq:eg}) can be reformulated as
\begin{equation}\label{eq:ex1}
\begin{split}
&\frac{R_{k_1,k_2;V_{\alpha,\bar{\alpha}},V_{-\alpha,-\bar{\alpha}}}}{R_{\mathbbm{1},\mathbbm{1},\mathbbm{1},\mathbbm{1}}}=\frac{1}{16 \pi^4 k^2}e^{-\frac{4\pi}{R}(k_1\tau_--k_2\tau_+)}I_{1234}\bar{I}_{1'2'3'4'};\\
&I_{1234}=\displaystyle \int_0^R dx_1dx_2 e^{\frac{2\pi i }{R}(k_1x_1-k_2x_2)}\braket{\partial \phi (\omega_1)\partial\phi(\omega_2)V_{\alpha}(\omega_3)V_{-\alpha}(\omega_4)}; \\
&\bar{I}_{1'2'3'4'}=\displaystyle \int_0^R dx_{1'}dx_{2'} e^{-\frac{2\pi i }{R}(k_1x_{1'}-k_2x_{2'})}\braket{\bar\partial \bar{\phi} (\bar{\omega}_{1'}) \bar\partial\bar{\phi}(\bar{\omega}_{2'})V_{\bar{\alpha}}(\bar{\omega}_3)V_{-\bar{\alpha}}(\bar{\omega}_4)}= I_{1234}^*.
\end{split}
\end{equation}
Applying the conformal transformation of the primary fields, we obtain for the four point correlation function involving both derivative and vertex operators the following
\begin{equation}
\begin{split}
I_{1234}&=-\left(\frac{\sin (\pi r)}{2 \sin (\pi r/2)}\right)^{\alpha^2} \displaystyle \int_0^R dx_1dx_2 e^{\frac{2\pi i }{R}(k_1x_1-k_2x_2)}\frac{dz}{d\omega}\Big|_{\omega=\omega_1}\frac{dz}{d\omega}\Big|_{\omega=\omega_2}\left(\frac{1}{z_{12}^2}+\frac{\alpha^2z_{34}^2}{z_{13}z_{23}z_{14}z_{24}} \right)\\
&=-\left(\frac{\sin (\pi r)}{2 \sin (\pi r/2)}\right)^{\alpha^2}\displaystyle \oint_{C_-}\oint_{C_+} dz_1dz_2e^{\frac{2\pi i }{R}(k_1x_1-k_2x_2)}\left(\frac{1}{z_{12}^2}+\frac{\alpha^2z_{34}^2}{z_{13}z_{23}z_{14}z_{24}} \right).
\end{split}
\end{equation}
 Let us notice that in evaluating the correlation function, any one of its term involves a single contraction of $\partial \phi(z_i)$ with one other operator: when contracted with another $\partial \phi(z_j)$, it gives a factor $-1/(z_i-z_j)^2$, while when contracted with a vertex operator $V_{\alpha_j}(z_j)$ gives $-i\alpha_j/(z_i-z_j)$. On the other hand, the terms of the correlation function can involve contractions of $V_{\alpha_i}(z_i)$ with multiple other operators.  When contracted with another $V_{\alpha_j}(z_j)$ it gives a contribution $(z_i-z_j)^{\alpha_i\alpha_j}$. These rules will be useful for treating the most general case in Section \ref{sec:generalcase}.

\begin{figure}
\centering
\subfigure
{\includegraphics[width=0.48\textwidth]{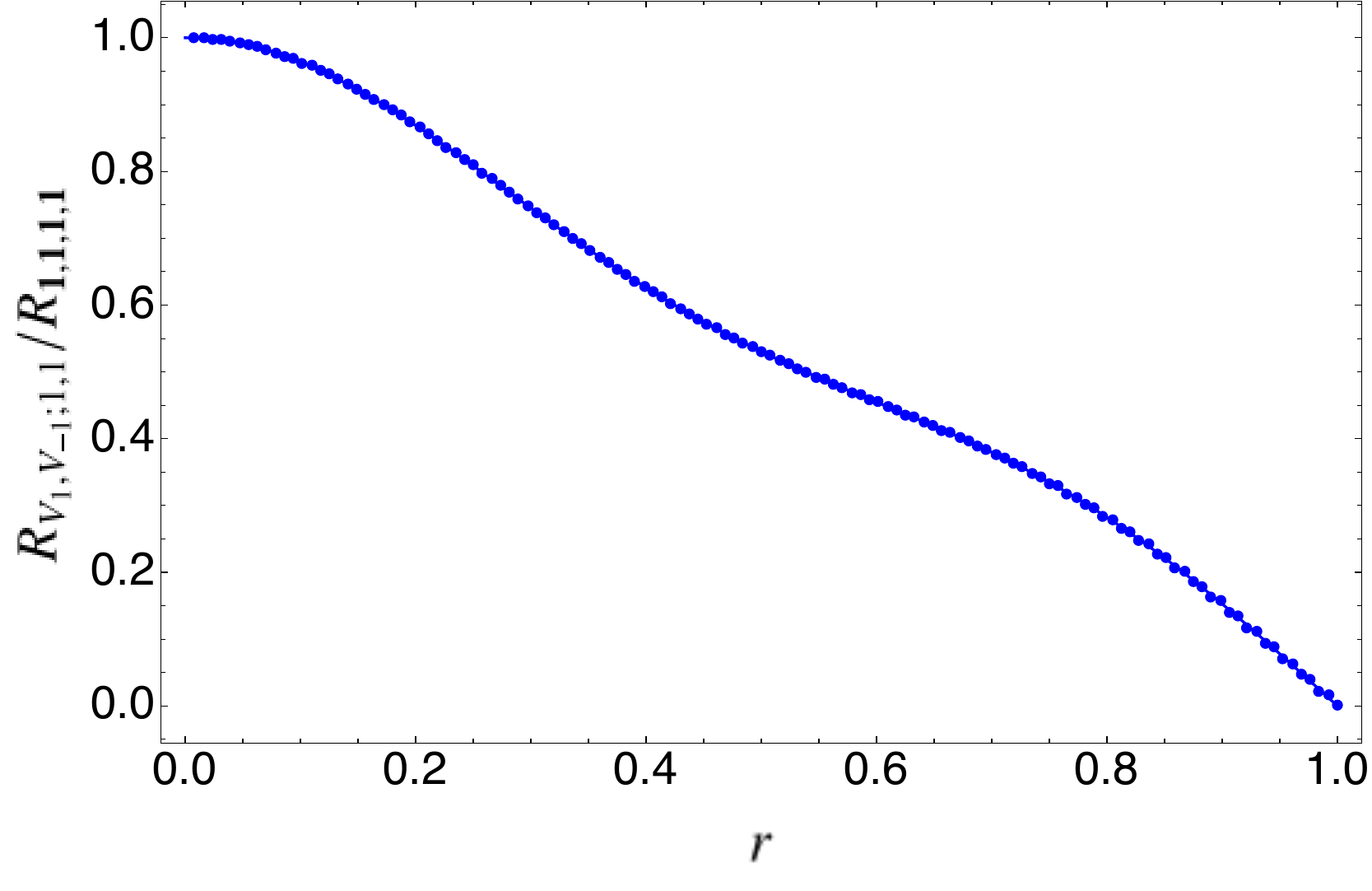}}
\subfigure
   {\includegraphics[width=0.48\textwidth]{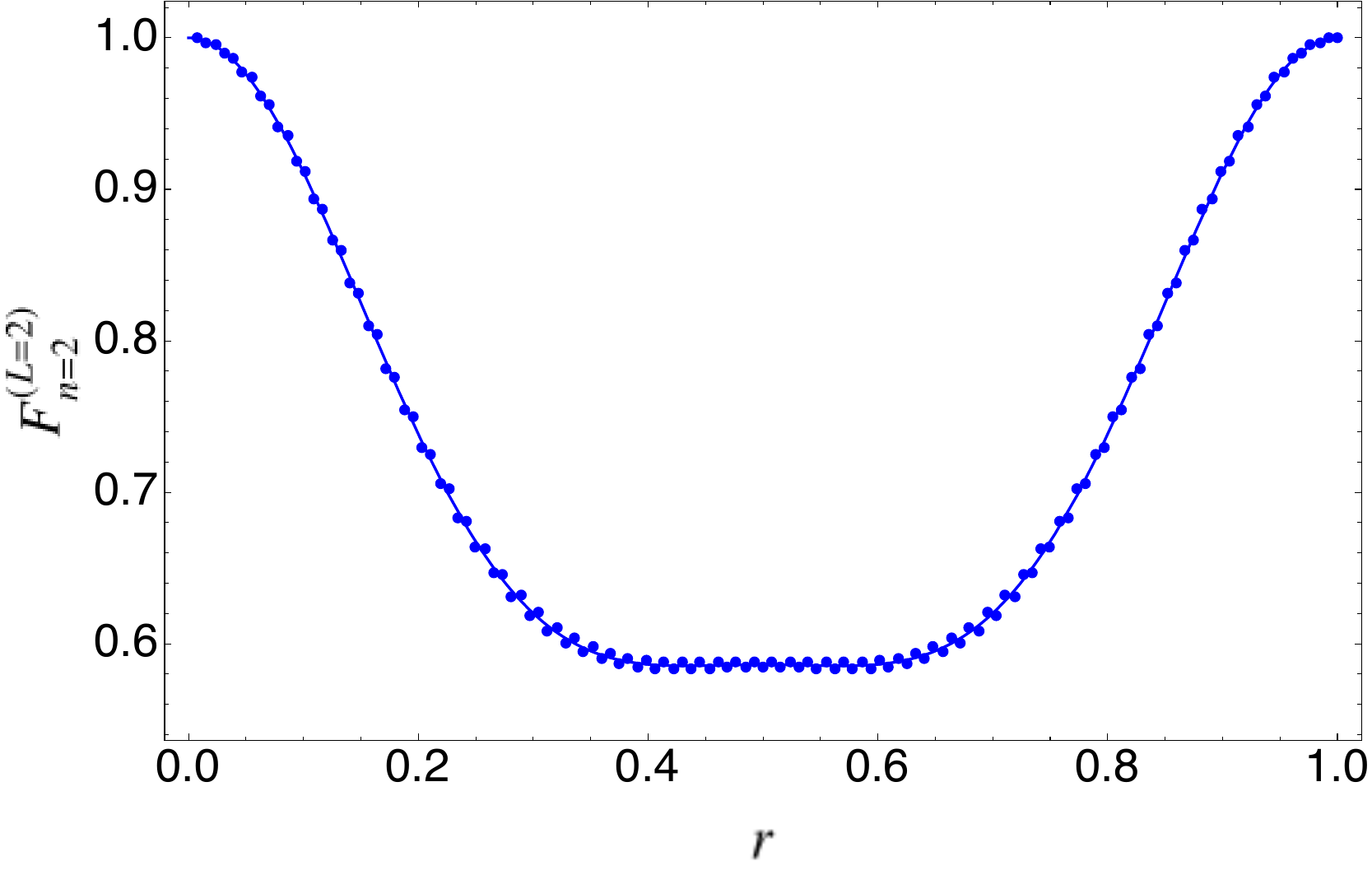}}
\caption{Further tests of the generalized R\'enyi entropies in the XX spin chains with $N=128$ spins.
Left panel: The quantity $\frac{R_{1,1;V_{1},V_{-1}}}{R_{\mathbbm{1},\mathbbm{1};\mathbbm{1},\mathbbm{1}}}$ as a function of $r=\ell/N$. 
The continuous line describes the analytical prediction in Eq. \eqref{eq:test2} while the symbols refer to the numerical data. 
Right panel: We test Eq. \eqref{eq:test3}, i.e. the excess of entropy for descendent states at level 2, as a function of $r=\ell/N$.}
\label{fig:p-vertex}
\end{figure}

Using Eqs.\,(\ref{eq:points2}), we get
\begin{equation}
\begin{split}
& I_{1234}=e^{\frac{2\pi}{R}(k_1\tau_- -k_2\tau_+)} \frac{4\pi^2 }{\Gamma(k_1)\Gamma(k_2)}S(r,\alpha,k_1,k_2),\\
& S(r,\alpha,k_1,k_2) \equiv  \left(\frac{\sin (\pi r)}{2 \sin (\pi r/2)}\right)^{\alpha^2} \\ &\times\partial_{z_1}^{k_1-1}\partial_{z_2}^{k_2-1}\left[\left(\frac{z_{1}^2-e^{-i\pi r}}{z_{1}+e^{i\pi r/2}} \right)^{k_1}\left(\frac{z_{2}^2-e^{i\pi r}}{z_{2}+e^{-i\pi r/2}} \right)^{k_2}\left(\frac{1}{z_{12}^2}+\frac{\alpha^2z_{34}^2}{z_{13}z_{23}z_{14}z_{24}} \right)\right]\Bigg|_{\substack{z_{1,3}=\pm e^{i\pi r/2}  \\ z_{2,4}=\pm e^{-i\pi r/2}}} , \\
&\frac{R_{k_1,k_2;V_{\alpha,\bar{\alpha}},V_{-\alpha,-\bar{\alpha}}}}{R_{\mathbbm{1},\mathbbm{1},\mathbbm{1},\mathbbm{1}}}=\frac{1}{\prod_{i=1}^2k_i \Gamma(k_i)^2}|S(r,\alpha,k_1,k_2)|^2.
\end{split}
\end{equation}
As $R_{0,0;1,1}/R_{\mathbbm{1},\mathbbm{1};\mathbbm{1},\mathbbm{1}}$ in Eq.\,(\ref{eq:results}) corresponds to $G(\rho_{GS}||\rho_{i\partial \phi})$, setting $k_1=k_2=1$, we obtain the relative entanglement entropy of the state generated by vertex operators $V_{\alpha}$ with respect to the state described by the chiral primary operator $i\partial \phi$, i.e.
\begin{multline}\label{eq:test2}
\frac{R_{1,1;V_{\alpha},V_{-\alpha}}}{R_{V_{\alpha},V_{-\alpha};V_{\alpha},V_{-\alpha}}}=
\frac{R_{1,1;V_{\alpha},V_{-\alpha}}}{R_{\mathbbm{1},\mathbbm{1};\mathbbm{1},\mathbbm{1}}}=\\ 
\left(\frac{\sin (\pi r)}{2 \sin (\pi r/2)}\right)^{\alpha^2}\frac{\sin^2(\pi r)}{4}  (\csc^2 (\pi r/2)+ \alpha^2\tan^2(\pi r/2)),
\end{multline}
where we used that the entropy of excitations corresponding to vertex operators coincides with the ground state entropy \cite{sierra}.
This expression has been checked against numerical computations in Fig.\,\ref{fig:p-vertex} for the chiral component of the bosonic field when $\alpha=1$. The CFT state generated by a vertex operator $V_{\alpha=1}\ket{0}$ corresponds in the XX chain to a particle-type excitation \cite{sierra}, as briefly reviewed in Appendix \ref{appendixTools}.

\subsection{A more complicated case: two chiral modes}
The previous two examples have shown that we can focus only on the chiral states, since the contributions of the chiral and anti-chiral parts to the GMSREs factorize.
Hence, let us consider an example in which there are states involving more than one chiral mode such as
\begin{equation}\label{eq:RTTTT}
R_{(1,1),(1,1);(1,1),(1,1)}=\mathrm{Tr}_A(\mathrm{Tr}_B(a_{-1}a_{-1}\ket{0}\bra{0}a_1a_1)\mathrm{Tr}_B(a_{-1}a_{-1}\ket{0}\bra{0} a_1a_1)).
\end{equation}
The Fourier expansion of the field in Eq. \eqref{eq:Fexp} leads to
\begin{align}
a_{-1}a_{-1}\ket{0} &= \frac{e^{-\frac{2\pi (\tau_1+\tau_2)}{R}}}{4\pi^2} \int dx_1dx_2e^{\frac{2\pi i  }{R}(x_1+x_2)}\partial_{\omega_1}\phi(\omega_1)\partial_{\omega_2}\phi(\omega_2)\ket{0};\\
\bra{0}a_1a_1 &= \frac{e^{\frac{2\pi (\tau_3+\tau_4)}{R}}}{4\pi^2} \int dx_3dx_4e^{-\frac{2\pi i  }{R}(x_3+x_4)}\bra{0}\partial_{\omega_3}\phi(\omega_3)\partial_{\omega_4}\phi(\omega_4).
\end{align}
Thus, the generalized entropy in Eq. \eqref{eq:RTTTT} reads
\begin{multline}\label{eq:2modes}
\frac{R_{(1,1),(1,1);(1,1),(1,1)}}{R_{\mathbbm{1},\mathbbm{1},\mathbbm{1},\mathbbm{1}}}=\frac{1}{\sqrt{2}^4}\frac{1}{(4\pi^2)^4}e^{-\frac{2\pi}{R}(\tau_1+\tau_2-\tau_3-\tau_4+\tau_5+\tau_6-\tau_7-\tau_8)}\\ \times \displaystyle \int_0^R dx_1\dots dx_8 e^{\frac{2\pi i }{R}(x_1+x_2-x_3-x_4+x_5+x_6-x_7-x_8)}\braket{\partial_{\omega_1}\phi(\omega_1) \dots \partial_{\omega_8}\phi(\omega_8)},
\end{multline}
where $\frac{1}{\sqrt{2}^4}$ normalizes the states in Eq. \eqref{eq:RTTTT}.
Once we perform the conformal mappings from the two-sheeted Riemann surface $\mathcal{R}_2$ to a single complex plane, the previous integral can be computed through the residue theorem. The points that the $w_i$'s are mapped onto are given by
\begin{equation}
y_1=e^{i\pi r/2}, \quad y_2=e^{-i\pi r/2}, \quad y_3=-e^{i\pi r/2}, \quad y_4=-e^{-i\pi r/2}.
\end{equation}
The difference with respect to the computations involving one single mode is that now the poles seen by the spatial integrals can arise both from $z_i=y_j$ but also from $z_i=z_j$. For future convenience it is useful to write down the correlation function $\braket{\partial_{z_1}\phi(z_1) \dots \partial_{z_8}\phi(z_8)}$ so that all occurrences when there are contractions involving $z_i$'s near the same point (one of the four $y_i$) are explicitly written, i.e.
\begin{multline}
\braket{\partial_{z_1}\phi(z_1) \dots \partial_{z_8}\phi(z_8)}=\braket{12345678}'+\frac{1}{z^2_{12}}\braket{345678}'+\frac{1}{z^2_{34}}\braket{125678}'+\frac{1}{z^2_{56}}\braket{123478}'\\+\frac{1}{z^2_{78}}\braket{123456}'+ \frac{1}{z^2_{12}z^2_{34}}\braket{5678}'+\frac{1}{z^2_{12}z^2_{56}}\braket{3478}'+\frac{1}{z^2_{12}z^2_{78}}\braket{3456}'+\frac{1}{z^2_{34}z^2_{56}}\braket{1278}' \\+\frac{1}{z^2_{34}z^2_{78}}\braket{1256}'+\frac{1}{z^2_{56}z^2_{78}}\braket{1234}'+\frac{1}{z^2_{12}z^2_{34}z^2_{56}z^2_{78}}.
\end{multline}
We use the prime, i.e. $\braket{\dots}'$, on the correlators to indicate that there are no
contractions involving (1, 2) or (3, 4) or (5, 6) or (7, 8). Hence, we can evaluate the primed correlation functions as
\begin{equation}
\begin{split}\label{eq:primed}
\braket{12345678}'=&\sum_{\substack{ijkl=1 \\ ijkl \, \mathrm{all}\,\mathrm{distinct}}}\frac{1}{2}\frac{1}{(y_i-y_j)^4}\frac{1}{(y_k-y_l)^4}+\\ &2\frac{1}{(y_i-y_j)^2}\frac{1}{(y_i-y_k)^2}\frac{1}{(y_j-y_l)^2}\frac{1}{(y_k-y_l)^2}; \\
\braket{1 \cdots \widehat{2i-1}\widehat{2i}\cdots 8}'=& 8\sum_{\substack{jkl \neq i \\ j<k<l}}\frac{1}{(y_j-y_k)^2}\frac{1}{(y_j-y_l)^2}\frac{1}{(y_k-y_l)^2}, \, i=1,2,3,4; \\ 
\braket{1 \cdots \widehat{2i-1}\widehat{2i}\cdots \widehat{2j-1}\widehat{2j} \cdots 8}'=& 2\sum_{\substack{k<l \\k,l \,\mathrm{distinct}\,\mathrm{from} \,i,j}}\frac{1}{(y_k-y_l)^4}, i<j, i,j=1,2,3,4.
\end{split}
\end{equation}
Let us now plug the results found in Eq. \eqref{eq:primed} in the integral in Eq. \eqref{eq:2modes}. We must evaluate integrals of the form:
\begin{equation}
\begin{split}
I_{12}&=\int_0^R dx_1dx_2e^{\frac{2\pi}{R}i(x_1+x_2)}\frac{1}{z^2_{12}}\partial_{\omega_1}z_1\partial_{\omega_2}z_2  \\
=& \oint_{C^1_{-}}dz_1\oint_{C^2_{-}}dz_2e^{\frac{2\pi}{R}(\tau_1+\tau_2)}e^{4\pi i v/R}\frac{1}{z_1-y_1}\frac{1}{z_2-y_1}\frac{1}{(z_1-z_2)^2}f(z_1,y_1)f(z_2,y_1),\\
I_{13}&= \int_0^R dx_1dx_3e^{\frac{2\pi}{R}i(x_1-x_3)}\frac{1}{z^2_{13}}\partial_{\omega_1}z_1\partial_{\omega_3}z_3  \\
&= \oint_{C^1_{-}}dz_1\oint_{C^3_{+}}dz_2e^{\frac{2\pi}{R}(\tau_1-\tau_3)}\frac{1}{z_1-y_1}\frac{1}{z_3-y_2}\frac{1}{(z_1-z_3)^2}f(z_1,y_1)f(z_3,y_2)\label{eq:I13},
\end{split}
\end{equation}
where 
\begin{equation}
f(z_i,y_j)=\frac{z_i^2-(y_j^*)^2}{z_i+y_j}.
\end{equation}
In order to do the first integral, we will take $\tau_1< \tau_2$ (without loss of generality), such that the contour $C^1_{-}$ is inside $C^2_{-}$, i.e. we can first perform the
integral $C^1_{-}$ followed by $C^2_{-}$ (similarly for the integral along $C^3_{+}$), with the result
\begin{equation}
\begin{split}
I_{12}=&-4\pi^2e^{2\pi (\tau_1+\tau_2)/R}e^{4\pi i v/R}f(y_1,y_1) \frac{1}{2}\partial^2_zf(z,y_1)|_{z=y_1}=-\frac{\pi^2}{y_1^2}e^{\frac{2\pi}{R}(\tau_1+\tau_2)}e^{4\pi i v/R} f(y_1,y_1)^2,\\
I_{13}=&4\pi^2\frac{1}{(y_1-y_2)^2}e^{\frac{2\pi}{R}(\tau_1-\tau_3)} f(y_1,y_1)f(y_2,y_2).
\end{split}
\end{equation}
The dependence on $\tau_i$'s cancel in Eq. \eqref{eq:2modes}, thus even though some $w_{i}$'s are mapped onto the same point once $\tau_i\to \pm \infty$, if we keep them finite, we can do the
integral and then take the limit, obtaining a finite result. 
We can then perform all the integrals along the lines discussed above and put everything together to get
\begin{multline}\label{eq:stressenergy}
\frac{R_{(1,1),(1,1);(1,1),(1,1)}}{R_{\mathbbm{1},\mathbbm{1},\mathbbm{1},\mathbbm{1}}}=\frac{1}{4}\Big[
 \frac{(7 + \cos(2 \pi r)) }{4096}(1606 + 335 \cos(2 \pi r) + 106 \cos(4 \pi r)+ \cos(6 \pi r)) \\ -\frac{\sin^6(\pi r)}{2}+\frac{(7 +\cos(2 \pi r))^2 \sin^4(\pi r)}{1024}+\frac{\sin^8(\pi r)}{256}\Big]=\\\frac{88123 + 37256 \cos(2 \pi r) + 4604 \cos(4 \pi r) + 
 1080 \cos(6 \pi r)+ 9 \cos(8 \pi x)}{131072}.
\end{multline}
We observe that this result coincides with the second R\'enyi entropy for the first descendant in the tower of a CFT with central charge $c=1$, i.e., the state associated with the stress-energy tensor $T = L_{-2} \mathbbm{1} $ ($\{L_p, p \in Z\}$ from the Virasoro algebra \cite{difrancesco}). 

\subsection{ GMSREs involving level 2 chiral states}
As a last example before we move to the most general case, we can follow the logic of the previous computations in order to evaluate all the second generalized mixed state entropies involving conformal level $L=2$ states, i.e. all the combinations involving the chiral states $a_{-1}a_{-1}\ket{0}$ and $a_{-2}\ket{0}$, without vertex operators. 
Working as before, we can easily find that 
\begin{equation}
\begin{split}\label{eq:co}
&\frac{R_{(1,1),(1,1);2,2}}{R_{\mathbbm{1},\mathbbm{1};\mathbbm{1},\mathbbm{1}}}= \frac{\cos^6(\pi r/2) (623 - 778 \cos(\pi r) + 376 \cos(2 \pi r) - 
   118 \cos(3 \pi r) + 25 \cos(4 \pi r))}{128},\\
   &\frac{R_{(1,1),2;2,(1,1)}}{R_{\mathbbm{1},\mathbbm{1};\mathbbm{1},\mathbbm{1}}}=\frac{\sin^6(\pi r/2) (623 - 778 \cos(\pi r) + 376 \cos(2 \pi r) - 
   118 \cos(3 \pi r) + 25 \cos(4 \pi r))}{128},\\
&\frac{R_{(1,1),2;(1,1),2}}{R_{\mathbbm{1},\mathbbm{1};\mathbbm{1},\mathbbm{1}}}=\frac{R_{2,(1,1);2,(1,1)}}{R_{\mathbbm{1},\mathbbm{1};\mathbbm{1},\mathbbm{1}}}= \frac{ 3 (71 + 9 \cos(2 \pi r) \sin^6(\pi  r)}{1024}, \\
&\frac{R_{2,2;2,2}}{R_{\mathbbm{1},\mathbbm{1};\mathbbm{1},\mathbbm{1}}}= \frac{ 22931 + 8072 \cos(2 \pi r) + 1628 \cos(4 \pi r) + 
  56 \cos(6 \pi x) + 81 \cos(8 \pi x)}{32768}.
\end{split}
\end{equation}
It is interesting to notice that the linear combination of the above quantities exactly reproduces the second R\'enyi entropy of the linear combinations of operators $T \pm L_{-1}\partial \phi$ found in \cite{palmai}, i.e.
\begin{multline}\label{eq:test3}
F^{(L=2)}_{n=2}\equiv 
\frac{1}{4}\Big( 2\frac{R_{(1,1),(1,1);2,2}}{R_{\mathbbm{1},\mathbbm{1};\mathbbm{1},\mathbbm{1}}}+2\frac{R_{(1,1),2;2,(1,1)}}{R_{\mathbbm{1},\mathbbm{1};\mathbbm{1},\mathbbm{1}}}+
2\frac{R_{(1,1),2;(1,1),2}}{R_{\mathbbm{1},\mathbbm{1};\mathbbm{1},\mathbbm{1}}}+
\frac{R_{(1,1),(1,1);(1,1),(1,1)}}{R_{\mathbbm{1},\mathbbm{1},\mathbbm{1},\mathbbm{1}}} 
+\frac{R_{2,2;2,2}}{R_{\mathbbm{1},\mathbbm{1};\mathbbm{1},\mathbbm{1}}} \Big)\\=
\frac{382659 + 106184 \cos(2\pi r) + 32924 \cos(4\pi r) + 2296 \cos(6\pi r) + 225 \cos(8\pi r) }{524288},
\end{multline} 
where we used that the generalized entropies are invariant under the exchange of replicas, $R_{i,j;i',j'}=R_{i',j';i,j}$.
We check this result against exact lattice computations in the right panel of Fig. \ref{fig:p-vertex}. In the XX chain the CFT state corresponds to two states with the same entropy, one made up of two holes and one particle and the other made up of one hole and two particles \cite{palmai}.

\section{GMSREs for arbitrary bosonic states}\label{sec:mgc}
In this section we report the results for the generalized R\'enyi entropies involving arbitrary bosonic CFT states.   To set the stage for the most general case, we first consider the case involving states with arbitrary mode content but no vertex operators.
  
\subsection{Generalized mixed state R\'enyi entropies without vertex operators}
Using what we have learnt from the previous simple examples, we now want to treat a much more general case, involving an arbitrary (even) number of modes but without vertex operators, i.e.
\begin{multline}
R_{k_1,\dots; \dots k_N}=A_1A_2A_3A_4\\ \mathrm{Tr}_A(\mathrm{Tr}_B(\prod_{i=1}^{N_1}a_{-k_i}\ket{0}\bra{0}\prod_{i=1}^{N_2}a_{k_{N_1+i}})\mathrm{Tr}_B(\prod_{i=1}^{N_3}a_{-k_{N_1+N_2+i}}\ket{0}\bra{0}\prod_{i=1}^{N_4}a_{k_{N_1+N_2+N_3+i}})),
\end{multline}
where $N=N_1+N_2+N_3+N_4$ is the total number of the modes.  Here $k_1, \dots k_N$ are all positive integers and the $A_j$'s denote the normalization of the states $a_{-k_{1}}^{n_{k_{1}}}\dots a_{-k_{M_j}}^{n_{k_{ M_{j}}}}\ket{0}$ which are given by
\begin{equation}
A_j=1/(\braket{0|\prod_{i=1}^{M_j}a_{k_i})\prod_{i=1}^{M_j}a_{-k_i}|0})^{1/2}=\frac{1}{\prod_{i=1}^{M_j} \sqrt{k_{i}^{n_{k_{i}} }(n_{k_{i} }!)}}.
\end{equation}
Using the integral representation of modes, we have to compute the following correlation function:
\begin{multline}
\langle \prod_{i=1}^{N_1}\partial_{\omega_i}\phi(\omega_i)\prod_{i=1}^{N_2}\partial_{\omega_i+N_1}\phi(\omega_{i+N_1})\prod_{i=1}^{N_3}\partial_{\omega_{i+N_1+N_2}}\phi(\omega_{i+N_1+N_2})\\ \prod_{i=1}^{N_4}\partial_{\omega_{i+N_1+N_2+N_3}}\phi(\omega_{i+N_1+N_2+N_3})\rangle=\prod_{i=1}^N\partial_{\omega_i}z_i \sum_{\substack{\sigma \in S_N  \\ \sigma_{2i}<\sigma_{2i+1}\\ \sigma_1<\sigma_3<\dots \sigma_{2N-1}}}\prod_{i=1}^{N/2}\frac{1}{(z_{\sigma_{2i-1}}-z_{\sigma_{2i}})^2},
\end{multline}
where the second line comes from the conformal mapping from the two-sheeted Riemann surface to the plane. We can recognize that the last sum over $\sigma \in S_N$, with $S_N$ the permutation group of the $N$ indices, can be compactly rewritten as $\mathrm{Hf} \left(\frac{1}{(z_i-z_j)^2}\right)_{1\leq i,j,\leq N}$, where $\mathrm{Hf }$ denotes the Hafnian of a matrix $B$
\begin{equation}
\mathrm{Hf(B)}=\frac{1}{2^{N/2} {N/2}!}\sum_{\sigma \in S_N}\prod_{i=1}^{N/2}B_{\sigma(2i-1),\sigma(2i)},
\end{equation}
an object which contains $(N-1)!!$ terms.
Let us now consider the needed integrations focusing on one of the  term in the above sum, i.e.
\begin{equation}
\prod_{i=1}^{N/2}\frac{1}{(z_{\sigma_{2i-1}}-z_{\sigma_{2i}})^2}.
\end{equation}
Its contribution to $R_{k_1,\dots;\dots k_N}/R_{\mathbbm{1},\mathbbm{1},\mathbbm{1},\mathbbm{1}}$ is 
\begin{equation}
\label{eq:p1}(-1)^{N_1+N_3}e^{2\pi i \frac{v}{R}(P_1+P_3-P_2-P_4)}\prod_{i=1}^{N/2}W(k_{\sigma_{2i-1}},k_{\sigma_{2i}},y_{\sigma_{2i-1}},y_{\sigma_{2i}}).
\end{equation}
The factor $(-1)^{N_1+N_3}$ comes from the $N$ contour integrals in total, $N_1+N_3$ that are clockwise and $N_2+N_4$ that are counterclockwise. 
Here $P_i$ (the total momentum of the state) are given by
\begin{equation}
P_i=\sum_{j=1}^{N_i}k_{j+\sum_{\ell=1}^{i-1}N_{\ell}},  \quad i=1\dots 4,
\end{equation}
and the $y_i$'s, $i = 1,\dots , N$, are defined as
\begin{equation}
y_i=\begin{cases}
e^{i\pi r/2}, & 1\leq i \leq N_1; \\
e^{-i\pi r/2}, & N_1+1 \leq i \leq N_1+N_2; \\
-e^{i\pi r/2}, & N_1+N_2+1 \leq i \leq N_1+N_2+N_3;\\
-e^{-i\pi r/2}, & N_1+N_2+N_3 +1\leq i \leq N_1+N_2+N_3+N_4.
\end{cases}
\end{equation}
The evaluation of  $W(k_i, k_j , y_i, y_j )$ in Eq. \eqref{eq:p1} (for conciseness, we use the subscript $(i,j)$ rather than $(\sigma_{2i-1},\sigma_{2i})$) leads to
\begin{equation}\label{eq:Wk's}
W(k_i, k_j , y_i, y_j )=\begin{cases}
\frac{1}{\Gamma(k_i)}\sum_{l=0}^{k_i-1} {{k_i-1}\choose{l}} \frac{\Gamma(k_i-l+1)}{\Gamma(k_i+k_j-l+1)}  (\partial_z^lf^{k_i})(z=y_i,y_i)\\ \qquad \times (\partial_z^{k_i+k_j-l}f^{k_j})(z=y_j,z=y_j), \quad &y_i=y_j; \\
\frac{1}{\Gamma(k_i)\Gamma(k_j)}\partial_{z_i}^{k_i-1}\partial_{z_j}^{k_j-1}\left(\frac{f^{k_i}(z_i,y_i)f^{k_j}(z_j,y_j)}{(z_i-z_j)^2} \right)\Big |_{\substack{z_i=y_i \\ z_j=y_j}} \quad &y_i=y_j;
\end{cases}
\end{equation}
 This first case above ($y_i=y_j$) can be obtained by exploiting the product rule for higher-order partial derivatives, i.e.
\begin{equation}
\partial^{\alpha}(hg)=\sum_{\beta + \gamma=\alpha}\frac{\alpha!}{\beta! \gamma!}(\partial^{\beta}h)(\partial^{\gamma}g),
\end{equation} 
whose proof can be done by induction. Using this rule for $h=f(z_i,y_i)$ and $g=1/(z_i-z_j)^2$, taking the derivative with respect to $z_i$, setting $z_i=y_i$, and then doing the integral in $z_j$, we obtain the first line of Eq. \eqref{eq:Wk's}. This corresponds to the assumption $\tau_i<\tau_j$, but we can explicitly check that nothing changes for $z_i \leftrightarrow z_j$.  We have applied this same logic in solving the integral $I_{12}$ in Eq. \eqref{eq:I13}. The second case of Eq. \eqref{eq:Wk's} is analogous to the solution of $I_{13}$: it can be obtained by simply applying the residue theorem since there are no contractions involving $z_i$'s ($i=1,\dots N$) near the same point (one of four $\pm e^{\pm i\pi r/2}$). 
To summarize, we have
\begin{equation}\label{eq:n2}
\frac{R_{k_1,\dots;\dots k_N}}{R_{\mathbbm{1},\mathbbm{1};\mathbbm{1},\mathbbm{1}}}=A_1A_2A_3A_4(-1)^{N_1+N_3}e^{2\pi i\frac{v}{R}(P_1+P_3-P_2-P_4)}\mathrm{Hf}\left( W\right),
\end{equation}
where $W$ is the matrix which enters in Eq. \eqref{eq:p1}.
This result can be extended to generalized R\'enyi entropies with index $n$. The total number of modes is now given by $N=\sum_{i=1}^{2n}N_i$, the set of $N$ points $y_k$ is defined by
\begin{equation}\label{eq:npoints}
y_k=\begin{cases} e^{i\pi \frac{j}{n}(r+m)} &\sum_{j=1}^{N_{m}}N_j<k<\sum_{j=1}^{N_{m+1}}N_j-1, \quad m\; \mathrm{even};\\
e^{i\pi \frac{j}{n}(-r+m-1)} &\sum_{j=1}^{N_{m}}N_j<k<\sum_{j=1}^{N_{m+1}} N_j-1, \quad m\; \mathrm{odd},
\end{cases}
\end{equation}
while the formal expression for $W(k_i,k_j,y_i,y_j)$ is the one defined in Eq. \eqref{eq:Wk's} with the following definition for $f$'s:
\begin{equation}
f(z_i,y_i)=\begin{cases}\dfrac{z_i^n-e^{i\pi r}}{\sum_{j=0}^{n-1}z_i^{n-j-1}e^{i\pi \frac{j}{n}(r+m)}}, &m\, \mathrm{even}; \\
\dfrac{z_i^n-e^{-i\pi r}}{\sum_{j=0}^{n-1}z_i^{n-j-1}e^{i\pi \frac{j}{n}(-r+m-1)}}, &m\, \mathrm{odd}.
\end{cases}
\end{equation}
Thus Eq. \eqref{eq:n2} becomes for generic $n$
\begin{equation}\label{eq:genn}
\frac{R_{k_1,\dots ;\dots k_N}}{R_{\mathbbm{1},\dots \mathbbm{1};\mathbbm{1},\dots \mathbbm{1}}}=A_1 \cdots A_{2n}(-1)^{\sum_{i\, \mathrm{odd}}N_i}e^{2\pi i\frac{v}{R}(\sum_{i\, \mathrm{odd}}P_i-\sum_{i\, \mathrm{even}}P_i)}\mathrm{Hf}\left( W\right).
\end{equation}

\subsection{Generalized mixed state R\'enyi entropies with vertex operators}\label{sec:generalcase}
The most general case involves states with arbitrary vertex operator content:
\begin{multline}\label{eq:Rvertex}
R^{\alpha_1,\alpha_2,\alpha_3,\alpha_4}_{k_1,\dots,;\dots k_N}=A^{\alpha_1}_1A^{\alpha_2}_2A^{\alpha_3}_3A^{\alpha_4}_4\\
\mathrm{Tr}_A(\mathrm{Tr}_B(\prod_{i=1}^{N_1}a_{-k_i}\ket{\alpha_1}\bra{\alpha_2}\prod_{i=1}^{N_2}a_{k_{N_1+i}})\mathrm{Tr}_B(\prod_{i=1}^{N_3}a_{-k_{N_1+N_2+i}}\ket{\alpha_3}\bra{\alpha_4}\prod_{i=1}^{N_4}a_{k_{N_1+N_2+N_3+i}}))
\end{multline}
where the states $\ket{\alpha_i}$ are defined via
\begin{equation}
\begin{split}
\ket{\alpha_{1,3}} &\equiv (\frac{R}{2\pi i}e^{\frac{2\pi i}{R}\omega_{\alpha_{1,3}}})^{\alpha^2_{1,3}/2}:e^{i\alpha_{1,3}\phi(\omega_{\alpha_{1,3}})}:\ket{0}\equiv c_{\alpha_{1,3}}:e^{i\alpha_{1,3}:\phi(\omega_{\alpha_{1,3}})}:\ket{0}, \quad \omega_{\alpha_{1,3}}=-i \infty \\
\bra{\alpha_{2,4}} &\equiv \bra{0} (\frac{R}{2\pi i}e^{-\frac{2\pi i}{R}\omega_{\alpha_{2,4}}})^{\alpha^2_{2,4}/2}:e^{i\alpha_{2,4}\phi(\omega_{\alpha_{2,4}})}:\equiv \bra{0} c_{\alpha_{2,4}}:e^{i\alpha_{2,4}\phi(\omega_{\alpha_{2,4}})}:, \quad \omega_{\alpha_{2,4}}=i\infty
\end{split}
\end{equation}
and the $A^{\alpha_j}_j$ are the normalization of the states where now the vacuum is replaced by the state created by the vertex operator, i.e.
\begin{equation}
A_j^{\alpha_j}=1/(\braket{\alpha_j|\prod_{i=1+N_1+\cdots N_{j-1}}^{N_j}a_{k_i}\prod_{i=1+N_1+\cdots N_{j-1}}^{N_j}a_{-k_i}|\alpha_j})^{1/2}.
\end{equation}
Eq. \eqref{eq:Rvertex} can be rewritten as 
\begin{multline}
\frac{R^{\alpha_1,\alpha_2,\alpha_3,\alpha_4}_{k_1,\dots,k_N}}{R_{\mathbbm{1},\mathbbm{1},\mathbbm{1},\mathbbm{1}}}=\frac{(-1)^N}{(2\pi)^N}e^{-\frac{2\pi}{R}\sum_{i=1}^N\sigma_ik_i\tau_i}A^{\alpha_1}_1A^{\alpha_2}_2A^{\alpha_3}_3A^{\alpha_4}_4\\
\times \int_0^R dx_1 dx_2 \dots dx_N e^{\frac{2\pi i}{R}\sigma_ik_ix_i}C(\omega_{\alpha_{1}},\omega_{\alpha_{2}},\omega_{\alpha_{3}},\omega_{\alpha_{4}},\omega_1,\dots, \omega_N),
\end{multline}
where $\sigma_i = \pm 1$ if the corresponding mode $a$ is a creation/annihilation operator, the factor $(-1)^N$ comes from representing each $a$ in terms of $\partial_{\omega}\phi(\omega)$, and \\ $C(\omega_{\alpha_{1}},\omega_{\alpha_{2}},\omega_{\alpha_{3}},\omega_{\alpha_{4}},\omega_1,\dots, \omega_N)$  is given by
\begin{multline}
C(\omega_{\alpha_{1}},\omega_{\alpha_{2}},\omega_{\alpha_{3}},\omega_{\alpha_{4}},\omega_1,\dots, \omega_N)=c_{\alpha_{1}}c_{\alpha_{2}}c_{\alpha_{3}}c_{\alpha_{4}}\langle \prod_{i=1}^{N_1}\partial_{\omega_i}\phi(\omega_i)e^{i\alpha_1\phi(\omega_{\alpha_1})}e^{i\alpha_2\phi(\omega_{\alpha_2})} \\ \times \prod_{i=1}^{N_2}\partial_{\omega_i+N_1}\phi(\omega_{i+N_1})\prod_{i=1}^{N_3}\partial_{\omega_{i+N_1+N_2}}\phi(\omega_{i+N_1+N_2})e^{i\alpha_3\phi(\omega_{\alpha_3})}e^{i\alpha_4\phi(\omega_{\alpha_4})}\\ \times \prod_{i=1}^{N_4}\partial_{\omega_{i+N_1+N_2+N_3}}\phi(\omega_{i+N_1+N_2+N_3})\rangle.
\end{multline} 
This correlation function can be evaluated by considering all the possible contractions among operators:  $e^{i\alpha_i\phi(\xi_i)}$ must be contracted to all other operators, with a contribution $(\xi_i-\xi_j)^{\alpha_i\alpha_j}$ when contracted with another $e^{i\alpha_j\phi(\xi_j)}$ and $-i\alpha_i/(\xi_i-z_j)$ when contracted with $\partial \phi(z_j)$. The operators $\partial \phi(z_i)$ must be contracted with one operator at a time, giving a factor $-1/(z_i-z_j)^2$ when contracted with another $\partial \phi(z_j)$. These rules were the same already used to evaluate the correlation function in Eq. \eqref{eq:ex1}.  

By summing up all these contributions, we get 
\begin{multline}
C(\omega_{\alpha_{1}},\omega_{\alpha_{2}},\omega_{\alpha_{3}},\omega_{\alpha_{4}},\omega_1,\dots, \omega_N)=M(\alpha_1,\alpha_2,\alpha_3,\alpha_4) \\
\times \prod_{i=1}^N \partial_{\omega_i}z_i\left[ \braket{\partial_{z_1}\phi\dots \partial_{z_N}\phi}+i\sum_{i}^4\sum_{i'}^N \frac{\alpha_i \alpha_j}{(\xi_i-z_{i'})} \braket{\prod_{l\neq i}^N  \partial_{z_l}\phi}\right.\\ \left.
-\sum_{i,j}^4\sum_{i'<j'}^N \frac{\alpha_i \alpha_j}{(\xi_i-z_{i'})(\xi_j-z_{j'})} \braket{\prod_{l\neq i,j}^N  \partial_{z_l}\phi}\right. \\ \left. + \sum_{i,j,k}^4\sum_{i'<j'<k'}^N \frac{\alpha_i \alpha_j\alpha_k }{(\xi_i-z_{i'})(\xi_j-z_{j'})(\xi_k-z_{k'})}  \braket{\prod_{l\neq i,j,k}^N  \partial_{z_l}\phi}+ \dots +i^N\sum_{i_1,i_2,\dots,i_N}^4 \prod_{j=1}^N\frac{\alpha_{i_j} }{(\xi_{i_j}-z_{{j}})}
\right] .
\end{multline}
The prefactor $M(\alpha_1,\alpha_2,\alpha_3,\alpha_4)$ encodes all information about the purely vertex operator part of the correlation function: 
\begin{multline}
M(\alpha_1,\alpha_2,\alpha_3,\alpha_4)=c_{\alpha_{1}}c_{\alpha_{2}}c_{\alpha_{3}}c_{\alpha_{4}}\braket{e^{i\alpha_1\phi(\omega_{\alpha_1})}e^{i\alpha_2\phi(\omega_{\alpha_2})}e^{i\alpha_3\phi(\omega_{\alpha_3})}e^{i\alpha_4\phi(\omega_{\alpha_4})} }\prod_{i=1}^4(\partial_{\omega_{\alpha_i}} z_i)^{\alpha_i^2/2}|_{z_i=y_i}\\
=\sin \left(\frac{\pi r}{2} \right)^{\alpha_1\alpha_2+\alpha_3\alpha_4}\cos \left(\frac{\pi r}{2} \right)^{\alpha_1\alpha_4+\alpha_2\alpha_3} \\ \times2^{-\bar{\alpha}\cdot \bar{\alpha}/2}\sin(\pi r)^{\bar{\alpha}\cdot \bar{\alpha}/2}  (e^{-i\pi r}e^{2\pi i v/R})^{(\alpha_1^2+\alpha_3^2-\alpha_2^2-\alpha_4^2)/2},
\end{multline}
\\ with $\bar{\alpha} = (\alpha_1,\alpha_2,\alpha_3,\alpha_4)$. When $\alpha_1=-\alpha_2=\alpha_3=-\alpha_4$, $M(\alpha_1,\alpha_2,\alpha_3,\alpha_4)=1$ as also found in Ref. \cite{sierra}.

Writing the correlation function in this way allows us to express $R^{\alpha_1,\alpha_2,\alpha_3,\alpha_4}_{k_1,\dots,k_N}$
in terms of R\'enyi entropies that do not involve the vertex operators, $R^{0,0,0,0}_{k_1,\dots,k_N}$, as follows:
\begin{multline}\label{eq:vv}
\frac{R^{\alpha_1,\alpha_2,\alpha_3,\alpha_4}_{k_1,\dots,k_N}}{R_{\mathbbm{1},\mathbbm{1},\mathbbm{1},\mathbbm{1}}}=M(\alpha_1,\alpha_2,\alpha_3,\alpha_4)\Big[ \frac{R^{0,0,0,0}_{k_1,\dots,k_N}}{R_{\mathbbm{1},\mathbbm{1},\mathbbm{1},\mathbbm{1}}} 
+\sum_{i=1}^N \frac{R^{0,0,0,0}_{k_1,\dots, \widehat{k_{i}} \dots k_N}}{R_{\mathbbm{1},\mathbbm{1};\mathbbm{1},\mathbbm{1}}}N(k_i)L_{k_i}(\bar{\alpha})\\+\sum_{i_1<i_2}^N \frac{R^{0,0,0,0}_{k_1,\dots, \widehat{k_{i_1}} \dots, \widehat{k_{i_2}},  \dots k_N}}{R_{\mathbbm{1},\mathbbm{1};\mathbbm{1},\mathbbm{1}}}N(k_{i_1},k_{i_2})L_{k_{i_1}}(\bar{\alpha})L_{k_{i_2}}(\bar{\alpha})+\dots+
\\\sum_{i_1<i_2,\dots < i_{N-2}}^N \frac{R^{0,0,0,0}_{k_1,\dots, \widehat{k_{i_1}} \dots, \widehat{k_{i_2}}, \dots\widehat{k_{i_{N-2}}} \dots k_N}}{R_{\mathbbm{1},\mathbbm{1};\mathbbm{1},\mathbbm{1}}}N(k_{i_1},k_{i_2},\dots,k_{i_{N-2}})L_{k_{i_1}}(\bar{\alpha})L_{k_{i_2}}(\bar{\alpha})\dots L_{k_{i_{N-2}}}(\bar{\alpha})  \\
+\prod_{i=1}^NL_{k_i}(\bar{\alpha}).
\Big ],
\end{multline}
where $\widehat{k_{i'} }$ indicates that the operator does not appear in the R\'enyi.
Here the coefficients $N(k_i)$ arise as normalizations of the states after some of the creation/annihilation operators have been removed since we are dealing with 
R\'enyi entropies involving only properly normalized states. So for example $N(k_i)$ for $i<N_1$ reads
\begin{equation}
    N(k_i)=\frac{(\braket{\alpha_1|\prod_{j\neq i}^{N_1}a_{k_j}\prod_{j\neq i}^{N_1}a_{-k_j}|\alpha_1})^{1/2}}{(\braket{\alpha_1|\prod_{j=1}^{N_1}a_{k_j}\prod_{j=1}^{N_1}a_{-k_j}|\alpha_1})^{1/2}}.
\end{equation}
The functions $L_i$ are defined as
\begin{equation}
\begin{split}
    L_{k_j}(\bar{\alpha})=&\sum_{i=1}^4 \alpha_i I_{ij}\\
    I_{ij}=&\frac{e^{-\frac{2\pi}{R} \tau_j \sigma_j k_j}}{2\pi i}\int_0^R dx_je^{\frac{2\pi}{R} i x_j \sigma_j k_j}\frac{1}{\xi_i-z_j}\partial_{\omega_j}z_j.
    \end{split}
\end{equation}
The integral above gives 
\begin{equation}\label{eq:Jij}
\begin{split}
I_{ij}=& \sigma_j e^{\frac{2\pi}{R}\sigma_j iv}J_{ij};\\
J_{ij}=& \begin{cases}
\frac{1}{\Gamma(k_j+1)}\partial_{z_j}^{k_j}f^{k_j}(z_j,k_j)  & i=j\\
\frac{1}{\Gamma(k_j)}\partial_{z_j}^{k_j-1}\frac{f^{k_j}(z_j,k_j) }{z_j-y_i} & i\neq j.
\end{cases}
\end{split}
\end{equation}
The terms like $e^{\frac{2\pi }{R}\sigma_j i v k_j}$ allow one to recover the factor $e^{\frac{2\pi}{R}(P_1+P_3-P_2-P_4)}$ in Eq. \eqref{eq:p1}.  
Eq. \eqref{eq:vv} is the main result of this work. 
Despite its involved expression, it provides an efficient way to compute any generalized mixed state R\'enyi entropy of a bosonic CFT involving arbitrary excited states.

The extension of Eq. \eqref{eq:vv} to the $n$-th generalized R\'enyi entropy can be performed by using the expression for $R_{k_1,\cdots,k_N}^{0,\cdots,0}$ in Eq. \eqref{eq:genn}, considering all the possible contractions between $e^{i\alpha_j\phi(z_j)}$ and $\partial \phi(z_i)$ and modifying the prefactor $M$ encoding the purely vertex contributions as
\begin{multline}
M(\alpha_1, \cdots, \alpha_{2n})=\prod_{i<j}|y_i-y_j|^{\alpha_i \alpha_j} \\ \times 2^{\bar{\alpha}\cdot \bar{\alpha}/2} n^{-\bar{\alpha}\cdot \bar{\alpha}/2}\sin(\pi r)^{\bar{\alpha}\cdot \bar{\alpha}/2}  (e^{-i\pi r}e^{2\pi i v/R})^{(\sum_{i\, \mathrm{odd}}\alpha_i^2-\sum_{ \,\mathrm{ even}}\alpha_i^2)/2},
\end{multline} 
where $y_i$'s are as defined in Eq. \eqref{eq:npoints}.

\section{Conclusions}\label{sec:concl}
In this work we considered the generalized mixed state R\'enyi entropy defined in Eq. \eqref{eq:renyin}. 
We developed a strategy to compute the aforementioned quantities for a bosonic conformal field theory using the representation of its Hilbert space in terms of massless modes and highest weight states created by vertex operators. 
This procedure allowed us to recover some known results about the entanglement of low-energy excitations represented by primary fields \cite{sierra,paola} 
and of descendant states \cite{palmai}. Although we provided explicit results only for a bosonic theory, our approach can be extended to a generic state of a CFT 
written as a product of modes of primary fields (for example the Ising CFT).

Our results represent the starting point to apply the truncated conformal space approach \cite{tcsa,James_2018} to the evaluation of the R\'enyi entropies in 
both equilibrium and non-equilibrium situations.  We use the results herein to describe the time evolution of the second R\'enyi entropy after a quench in the sine-Gordon model \cite{mck-21}. 

A straightforward use of our results is the computation of entanglement measures involving two CFT states such as the relative entropy \cite{lashkari2014,paola} and the trace distance \cite{zrc-19,zrc-19b}.
Another relatively simple extension concerns the resolution of entanglement in systems endowed with an abelian symmetry \cite{goldstein,chen-21}.  Work in this direction is already in progress. 
A more challenging idea would be to derive similar non-diagonal objects for the negativity \cite{CCT,cct-13} which is a good measure of entanglement in mixed states obtained by constructing non-diagonal partially transposed reduced density matrices. 

\section*{Acknowledgments}
We thank Jiaju Zhang and Luca Capizzi for useful discussions. PC and SM acknowledge support from ERC under Consolidator grant number 771536 (NEMO).  R.M.K. was supported by the U.S. Department of Energy, Office of Basic Energy Sciences, under Contract No. DE-AC02-98CH10886.

\begin{appendices}
\section{Short-interval expansions}
In this appendix we use the operator product expansion (OPE) of twist fields  \cite{cct-11} to find the short-distance behavior of the second generalized R\'enyi entropy 
involving arbitrary (chiral) excited states. 

In the replica approach, the moments of the RDM, $\mathrm{Tr}\rho_A^n$, are evaluated for any (1+1)-dimensional quantum field theory as partition functions over the $n$-sheeted Riemann surface, $\mathcal{R}_n$, in which the $n$ sheets (replicas) are cyclically joined along the subsystem $A=[0,\ell]$. This partition function can be rewritten in terms of the correlator of twist fields, $\mathcal{T}_n, \mathcal{\tilde{T}}_n$, implementing twisted boundary conditions \cite{cc-09}. 

Let us consider the generalized reduced density matrix of the 2-replicated theory $\rho_A \equiv \mathrm{Tr}_B(\ket{\Upsilon_1 \Upsilon_3}\bra{\Upsilon_2 \Upsilon_4})$, 
defined on two copies of the original CFT. We denote by $w$ the coordinates on the physical manifold, the cylinder, and by $z$ the coordinate on the complex plane 
$\mathbb{C}$ after the conformal mapping $z=e^{2 i \pi w/R}$. Therefore $0$ and $\ell$ are respectively mapped into $1$ and $e^{2\pi i r}$.
The R\'enyi $n=2$ entropy of the ground state is 
\begin{equation}
R_{\mathbbm{1} ,\mathbbm{1} ;\mathbbm{1} ,\mathbbm{1} }=c_2 \braket{\mathcal{T}_2(0)\mathcal{\tilde{T}}_2(\ell)}_{1^{\otimes 2}},
\end{equation}
where $\braket{\cdot}_{1^{\otimes 2}}$ denotes that the correlation function is taken on a theory described by $\rho_A \equiv \mathrm{Tr}_B(\ket{\mathbbm{1} \mathbbm{1}}\bra{\mathbbm{1} \mathbbm{1}})$.  Within these notations, we want to compute 
\begin{equation}
\frac{R_{k_1\dots k_N}}{R_{\mathbbm{1} ,\mathbbm{1} ;\mathbbm{1} ,\mathbbm{1} }}=\frac{\braket{\mathcal{T}_2(0)\mathcal{\tilde{T}}_2(\ell)}_{\rho_A}}{\braket{\mathcal{T}_2(0)\mathcal{\tilde{T}}_2(\ell)}_{1^{\otimes 2}}}=\frac{\braket{\mathcal{T}_2(1)\mathcal{\tilde{T}}_2(e^{2\pi i x})}_{\rho_A}}{\braket{\mathcal{T}_2(1)\mathcal{\tilde{T}}_2(e^{2\pi i x})}_{1^{\otimes 2}}}.
\end{equation}
Since $\mathcal{\tilde{T}}_2=\mathcal{T}^{\dagger}_2$, in the fusion between $\mathcal{T}_2$ and $\mathcal{\tilde{T}}_2$ the identity is present, i.e.
\begin{equation}
[\mathcal{T}_2]\times [\mathcal{\tilde{T}}_2] \to \mathbbm{1} ,
\end{equation} 
and then all the descendants of the conformal tower of the identity are generated in the OPE \cite{cct-11}. 
Hence, the leading term is 1 when $\braket{\mathbbm{1} }_{\rho_A} \neq 0$, i.e. when $\Upsilon_1=\Upsilon_2$ and $\Upsilon_3=\Upsilon_4$, 
since we deal with a set of orthonormal states. 
The next to leading order term is given by the stress energy tensor appearing in the fusion between $\mathcal{T}_2$ and $\mathcal{\tilde{T}}_2$, i.e. \cite{cct-11}
\begin{equation}\label{eq:TT}
\begin{split}
\frac{\braket{\mathcal{T}_2(1)\mathcal{\tilde{T}}_2(e^{2\pi i r})}_{\rho_A}}{\braket{\mathcal{T}_2(1)\mathcal{\tilde{T}}_2(e^{2\pi i r})}_{1^{\otimes 2}}}&\simeq 1+h_{\mathcal{T}_2}(2\pi i r)^2\braket{\Upsilon_1\Upsilon_3 |\sum_{j=1,2}T_j(1)|\Upsilon_1\Upsilon_3}+\dots\\
& \simeq 1-\frac{(\pi r)^2}{4} (h_1+h_3),
\end{split}
\end{equation}
where $\sum_{j=1,2}T_j(1)$ is the the sum of the stress tensors of each replica, $h_{\mathcal{T}_2}=1/16$ \cite{cc-04} and $h_1,h_3$ are the conformal dimensions of the 
${\Upsilon_1}$ and ${\Upsilon_3}$, respectively. 
This result can be checked by considering the examples reported in the main text. In particular, the chiral parts of the entropies explicitly studied in Section \ref{sec3} read at short distance
\begin{equation}
\begin{split}
&\frac{R_{1,1;1,1}}{R_{\mathbbm{1},\mathbbm{1};\mathbbm{1},\mathbbm{1}}}=1-\frac{\pi^2r^2}{2}+O(r^4)\quad (h_1,h_3)=(1,1);\\
&\frac{R_{1,1;0,0}}{R_{\mathbbm{1},\mathbbm{1};\mathbbm{1},\mathbbm{1}}}=1-\frac{\pi^2r^2}{4}+O(r^4);\quad (h_1,h_3)=(1,0)\\
&\frac{R_{1,1;V_{\alpha},V_{-\alpha}}}{R_{V_{\alpha},V_{-\alpha};V_{\alpha},V_{-\alpha}}}=1-\frac{\pi^2r^2}{4}(1+\frac{\alpha^2}{2})+O(r^4);\quad (h_1,h_3)=(1,\frac{\alpha^2}{2})\\
&\frac{R_{(1,1),(1,1);(1,1),(1,1)}}{R_{\mathbbm{1},\mathbbm{1},\mathbbm{1},\mathbbm{1}}}=1-\pi^2r^2 \quad (h_1,h_3)=(2,2)\\
&\frac{R_{2,2;2,2}}{R_{\mathbbm{1},\mathbbm{1},\mathbbm{1},\mathbbm{1}}}=1-\pi^2r^2 \quad (h_1,h_3)=(2,2)\\
&\frac{R_{(1,1),(1,1);2,2}}{R_{\mathbbm{1},\mathbbm{1},\mathbbm{1},\mathbbm{1}}}=1-\pi^2r^2 \quad (h_1,h_3)=(2,2)\\
\end{split}
\end{equation}
This is in agreement with the result found in Eq. \eqref{eq:TT}.

Let us now discuss the effect of the operators that are not in the tower of the identity. 
Let us assume that a primary operator  ${\cal O}\neq \mathbbm{1}$ is present in the OPE between $\mathcal{T}_2$ and $\tilde{\mathcal{T}}_2$.
When $\Upsilon_1=\Upsilon_2$, $\Upsilon_3=\Upsilon_4$, the one point function of ${\cal O}$ vanishes. 
Instead the two (or $n$)-point function of ${\cal O}$ contributes as $o(r^2)$ and so it is subleading compared to the stress energy tensor.

Let us move to the the off diagonal terms and consider the contribution of the derivative operator $i\partial \phi$ in the OPE between $\mathcal{T}_2$ and $\tilde{\mathcal{T}}_2$.
When $\Upsilon_1$ and $\Upsilon_2$ (or, equivalently,  $\Upsilon_3$ and $\Upsilon_4$) are descendants of the derivative operator whose level differs exactly by 1, 
we have a non-vanishing contribution at $O(r^2)$ (not $1+O(r^2)$ but only $O(r^2)$). 
In terms of modes, $\Upsilon_1$ and $\Upsilon_3$ (equivalently $\Upsilon_2$ and $\Upsilon_4$) must differ by one mode $a_{-1}$.
For this $\rho_A$, the one-point correlation function of $i\partial \phi$ does not vanish and we have \cite{jiaju} 
\begin{equation}\label{eq:der}
\frac{\braket{\mathcal{T}_2(1)\mathcal{\tilde{T}}_2(e^{2\pi i r})}_{\rho_A}}{\braket{\mathcal{T}_2(1)\mathcal{\tilde{T}}_2(e^{2\pi i r})}_{1^{\otimes 2}}}\simeq \frac{(\pi r)^2}{4}.
\end{equation}
This is confirmed by the short-distance expansion of the generalised entropies (for the chiral part)
\begin{equation}
\begin{split}
&\frac{R_{0,1;1,0}}{R_{\mathbbm{1},\mathbbm{1};\mathbbm{1},\mathbbm{1}}}=\frac{R_{1,0;0,1}}{R_{\mathbbm{1},\mathbbm{1};\mathbbm{1},\mathbbm{1}}}=\frac{\pi^2 r^2 }{4}+O(r^4).
\end{split}
\end{equation}

The short distance behavior described by Eqs. \eqref{eq:TT} and \eqref{eq:der} signals that, for small $r$, the most relevant terms correspond to equal states in each replica or 
which differ by at most a single mode. 

\section{Numerical tools} \label{appendixTools}

This section describes how the numerical data reported in the main text have been obtained following Refs. \cite{vidal1,peschel2001,peschel2003,pe-09}. 

\subsection{Correlation matrices of excited states in the XX spin chain}

We consider the XX model for $N$ spin-$1/2$ particles with periodic boundary conditions
\begin{equation}
\label{eq:modif12}
H_{XX}=-\dfrac{1}{4}\sum_{j=1}^{N} \left( \sigma^x_j \sigma^x_{j+1}+\sigma^y_j \sigma_{j+1}^y \right),
\end{equation}
where $\sigma^{\alpha}$ are the Pauli matrices. By performing a Jordan-Wigner transformation, Eq. \eqref{eq:modif12} can be expressed in terms of spinless fermions, $\{ c^{\dagger}_m,c_n \}=\delta_{mn}$. 
We are interested in the generic correlation matrix of an eigenstate of the XX spin chain specified by a set of occupied momenta $\Omega=\{k_i\}$. To achieve this goal, we first introduce the $2N$ spatial Majorana modes defined as
\begin{equation}
\label{eq:majorana}
\begin{cases}
a_{2m-1}=c^{\dagger}_m+c_m \\
a_{2m}=i(c^{\dagger}_m-c_m).
\end{cases}
\end{equation}
The correlation matrix $\Gamma^{\Omega}_{mn}$ written in terms of Majorana operators is
\begin{equation}
\label{eq:gammamajo}
\Gamma^{\Omega}_{mn}=\braket{a_ma_n}_{\Omega}-\delta_{mn},
\end{equation} 
where $\braket{\cdots}_{\Omega}$ denotes the expectation value on the state labelled by $\Omega$. It takes the form
\begin{equation}
\label{eq:matrix}
\Gamma^{\Omega}= \left(\begin{matrix} 
\Pi_0 & \Pi_1 & \cdots &\Pi_{N-1} \\
\Pi_{-1} & \Pi_0 & \dots  \\
\cdots & \cdots & \ddots & \vdots \\
\Pi_{-N+1}  & \cdots & \qquad &\Pi_{0} \\ 
\end{matrix}\right), \qquad
\Pi_m=\left(\begin{matrix} 
g_m^{(1)} & g_m^{(2)} \\
-g_{-m}^{(2)} & g_m^{(1)} \\
\end{matrix}\right).
\end{equation}
From \eqref{eq:gammamajo}, the elements of $\Gamma^{\Omega}_m$ are given by
\begin{equation}
\label{eq:gmn}
\begin{cases}
&g^{(1)}_{m-n}=\braket{a_{2m}a_{2n}}_{\Omega}-\delta_{mn}=\braket{a_{2m-1}a_{2n-1}}_{\Omega}-\delta_{mn} \\
&g^{(2)}_{m-n}=\braket{a_{2m-1}a_{2n}}_{\Omega}.
\end{cases}
\end{equation}
These quantities can be expressed by evaluating the correlation functions of Majorana fermions $a_m$ in terms of the fermionic ones $c_m$. The trick is performing a Fourier transform and using the correlation function of the free fermionic variables, so that we find
\begin{equation}\label{eq:gmn1}
\begin{cases}
&g^{(1)}_{m-n}=\braket{c^{\dagger}_mc_n}_{\Omega}+\braket{c_mc^{\dagger}_n}_{\Omega} -\delta_{mn}=\frac{1}{N} \left[\sum_{k \in \Omega}e^{-i\frac{\pi k }{N}(m-n)}+\sum_{k \not \in \Omega}e^{i\frac{\pi k }{N}(m-n)}  \right] -\delta_{mn} \\
&g^{(2)}_{m-n}=-i\braket{c^{\dagger}_mc_n}_{\Omega}+i\braket{c_mc^{\dagger}_n}_{\Omega}=\frac{i}{N} \left[-\sum_{k \in \Omega}e^{-i\frac{\pi k }{N}(m-n)}+\sum_{k \not \in \Omega}e^{i\frac{\pi k }{N}(m-n)}  \right].
\end{cases}
\end{equation}

We are interested in the computation of entanglement between a spatial partition containing $l$ sites of the system and the rest of the system. The reduced density matrix is encoded in a $2l$-block of matrix $\Gamma^{\Omega}$, $\Gamma^{\Omega}_l$ \cite{vidal1}:
\begin{equation}
\label{eq:matrix1}
\Gamma^{\Omega}_l= \left(\begin{matrix} 

\Pi_0 & \Pi_1 & \cdots &\Pi_{l-1} \\
\Pi_{-1} & \Pi_0 & \dots  \\
\cdots & \cdots & \ddots & \vdots \\
\Pi_{-l+1}  & \cdots & \qquad &\Pi_{0} \\ 
\end{matrix}\right).
\end{equation}
Let $V \in SO(2l)$ be such that it brings $\Gamma^{\Omega}_l$ into a block-diagonal form $\tilde{\Gamma}^{\Omega}_l=V\Gamma^{\Omega}_l V^T$,
\begin{equation}
\label{eq:matrix2}
\tilde{\Gamma}^{\Omega}_l =\bigoplus_{r=1}^l  \left(\begin{matrix} 
0 & \nu_r \\
-\nu_r & 0   \\ 
\end{matrix}\right).
\end{equation}
By inverting \eqref{eq:majorana}, the corresponding true fermions are given by $c_m=\dfrac{a_{2m-1}+ia_{2m}}{2}$, which fulfill
\begin{equation}
\label{eq:correlationn}
\braket{c_mc_n}=0, \qquad \braket{c^{\dagger}_mc_n}=\delta_{mn}\dfrac{1+\nu_m}{2}.
\end{equation}
Since the $l$ fermionic modes are uncorrelated , the reduced correlation matrix of the block is a product 
\begin{equation}
\label{eq:productstate}
\rho_A=\rho_1 \otimes \cdots \otimes \rho_l,
\end{equation}
whose eigenvalue for each mode is given by $\zeta_k=\dfrac{1+\nu_k}{2}$ if it is occupied and $1-\zeta_k=\dfrac{1-\nu_k}{2}$ if it is unoccupied.
Once the eigenvalues of the reduced density matrix $\rho_A$ are known, the second R\'enyi entropy can be read off
\begin{equation}
\label{eq:entropyR}
S_l^{(2)}=-\sum_j \log \left[ \left( \dfrac{1+\nu_j}{2} \right)^2+ \left( \dfrac{1-\nu_j}{2} \right)^2 \right]=-\sum_j\log(1/2+\nu_j^2/2).
\end{equation}

We can also identify the low-lying energy eigenstates in the spin chain with the corresponding ones in the bosonic CFT used in the main text. For the XX spin chain, this identification has been discussed, for example, in Ref. \cite{sierra}. If $N$ is an even integer and multiple of 4, the identification of states in the spin chain for the derivative and the vertex operator are, respectively, 
\begin{equation}
\begin{split}
c_{\frac{N}{4}-\frac{1}{2}} c^{\dagger}_{\frac{N}{4}+\frac{1}{2}} \ket{0} &\leftrightarrow i\partial \phi,\\
c^{\dagger}_{\frac{N}{4}+\frac{1}{2}}  \ket{0} &\leftrightarrow V_{1,0},
\end{split}
\end{equation}
where  $\ket{0}$ is the ground state of the half-filled fermionic model.

\subsection{Product of gaussian operators} \label{appendixGamma}
When computing the second R\'enyi relative entropies between the reduced density matrices of two different eigenstates, we need to evaluate objects of the following form
\begin{equation}
\label{eq:goal}
\mathrm{Tr}(\rho_1 \rho_0).
\end{equation}
If $\rho_1$ and $\rho_0$ do not commute, they cannot be simultaneously diagonalised. 
However, it is possible to use the composition properties of gaussian density matrices: let $\rho [\Gamma]$ denote a gaussian density matrix characterized  by its correlation function $\Gamma$, written in terms of Majorana fermions. 
The composition of correlation matrices is indicated by $\Gamma \times \Gamma'$ and it is implicitly defined by
\begin{equation}
\label{eq:product}
\rho[\Gamma]\rho[\Gamma']= \mathrm{Tr}[\rho[\Gamma]\rho[\Gamma']]\rho[\Gamma \times \Gamma'],
\end{equation}
leading to \cite{fc-10}
\begin{equation}
\label{eq:productrule}
\Gamma \times \Gamma'=1-(1-\Gamma')\dfrac{1}{1+\Gamma \Gamma'}(1-\Gamma).
\end{equation}
The trace of two fermionic operators can be computed as
 \begin{equation}
 \label{eq:productrule1}
 \{ \Gamma, \Gamma' \} \equiv \mathrm{Tr}(\rho_{\Gamma}\rho_{\Gamma'})=\prod_{\mu \in \mathrm{Spectrum} [\Gamma \Gamma']/2}\dfrac{1+\mu}{2},
 \end{equation}
namely  it is the product of the eigenvalues of ($1 + \Gamma \Gamma')/2$ with halved degeneracy.

 \section{Values of the GMSREs for $r=1/2$}
 
In this last appendix we review the case of an equal bipartition (i.e $r=1/2$) in which the explicit expression for the generalized R\'enyi entropies simplifies.  These results can be useful to implement an algorithm which computes these object for an arbitrary sequence of modes.

We can derive a generalized form for $R_{k,k;0,0}(r)$, since $S(r,k)\equiv S(r,k,k,0,0)$ behaves as
\begin{equation}
S(r,k)=\sin^2(\pi r)\Big(c_0+\sum_{n=1}^{k-1}c_n\cos (2n r) \Big).
\end{equation}
When $r = 1/2$, we have for general $k$
\begin{equation}
S(1/2,k)= 
\begin{cases}
  - \dfrac{(k-1)!!^4}{4k} \quad & k \mbox{    even,  } \\
      -\dfrac{(k-2)!!^4k}{4} \quad & k \mbox{    odd.  } \\
\end{cases}
\end{equation}
A simplification arises also for $R_{k,0;0,k}(r)$, since $S(r,k)\equiv S(r,k,0,0,k)$ behaves as
\begin{equation}
S(r,k)=\tilde{c}_0+\sum_{n=1}^{k}\tilde{c}_n\cos \Big( \frac{(2n-1) r\pi}{2}\Big) , \qquad S(1/2,k)= - \frac{(k-1)!k!}{2}. 
\end{equation}
The above examples suggest that a possible compact expression for the generalised entropies can be found when $r=1/2$.
Here we report them. 
First of all let us notice that
\begin{multline*} 
W(k_i,k_j,e^{i\pi/4},-e^{-\mathrm{i}\pi/4})= W(k_i,k_j,-e^{i\pi/4},e^{-\mathrm{i}\pi/4})=\\ -W(k_i,k_j,-e^{i\pi/4},-e^{-\mathrm{i}\pi/4})=- W(k_i,k_j,e^{i\pi/4},e^{-\mathrm{i}\pi/4})
\end{multline*}  
and $W(k_i,k_j,e^{i\pi/4},-e^{-\mathrm{i}\pi/4})$ reads
\begin{multline}\label{eq:first}
 W(k_i,k_j,e^{i\pi/4},-e^{-\mathrm{i}\pi/4})= \\
\begin{cases}
   \mathrm{i}\frac{(k_i-1)!!^2(k_j)!!^2}{2(k_i-k_j)\Gamma(k_i)\Gamma(k_j+1)}= \mathrm{i}\frac{(k_i-1)!!(k_j)!!}{2(k_i-k_j)(k_i-2)!!(k_j-1)!!} &\quad \mbox{if  } k_i \mod 2 =0, k_j \mod 2 =1\\
    \mathrm{i}\frac{(k_j-1)!!^2(k_i)!!^2}{2(k_i-k_j)\Gamma(k_i+1)\Gamma(k_j)}=\mathrm{i}\frac{(k_j-1)!!(k_i)!!}{2(k_i-k_j)(k_j-2)!!(k_i-1)!!} &\quad \mbox{if  } k_j \mod 2 =0, k_i\mod 2 =1\\
  -\frac{(k_i-1)!(k_i)!}{2\Gamma(k_i)\Gamma(k_i)}=-\frac{k_i}{2} &\quad \mbox{if  } k_i= k_j \\
       0 &\quad \mbox{otherwise.  } \\
\end{cases}
\end{multline}
For $W(k_i,k_j,e^{\mathrm{i}\pi/4},-e^{\mathrm{i}\pi/4})= W(k_i,k_j,e^{-\mathrm{i}\pi/4},-e^{-\mathrm{i}\pi/4})$, we find that
\begin{multline}
 W(k_i,k_j,e^{\mathrm{i}\pi/4},-e^{\mathrm{i}\pi/4})= \\
\begin{cases}
   \frac{-(k_i)!!^2(k_j)!!^2}{2(k_i+k_j)\Gamma(k_i+1)\Gamma(k_j+1)} = \frac{-(k_i)!!(k_j)!!}{2(k_i+k_j)(k_i-1)!!(k_j-1)!!} &  \mbox{if  } k_i \mod 2 =1, k_j \mod 2 =1\\
   \frac{-(k_i-1)!!^2(k_j-1)!!^2}{2(k_i+k_j)\Gamma(k_i)\Gamma(k_j)}= \frac{-(k_i-1)!!(k_j-1)!!}{2(k_i+k_j)(k_i-2)!!(k_j-2)!!}  &  \mbox{if  } k_i \mod 2 =0, k_j \mod 2 =0\\
       0 & \mbox{otherwise.  } \\
\end{cases}
\end{multline}
Finally, for $y_i=y_j$, we notice that 
\begin{multline*} 
W(k_i,k_j,e^{\mathrm{i}\pi/4},e^{\mathrm{i}\pi/4})= W(k_i,k_j,e^{-\mathrm{i}\pi/4},e^{-\mathrm{i}\pi/4}) \\= W(k_i,k_j,-e^{\mathrm{i}\pi/4},-e^{\mathrm{i}\pi/4})= W(k_i,k_j,-e^{-\mathrm{i}\pi/4},-e^{-\mathrm{i}\pi/4})
\end{multline*}
 and
\begin{multline}\label{eq:last}
 W(k_i,k_j,e^{\mathrm{i}\pi/4},e^{\mathrm{i}\pi/4})= \\
\begin{cases}
   \frac{(k_i)!!^2(k_j)!!^2}{2(k_i+k_j)\Gamma(k_i+1)\Gamma(k_j+1)} =  \frac{(k_i)!!(k_j)!!}{2(k_i+k_j)(k_i-1)!!(k_j-1)!!} &  \mbox{if  } k_i \mod 2 =1, k_j \mod 2 =1\\
   \frac{(k_i-1)!!^2(k_j-1)!!^2}{2(k_i+k_j)\Gamma(k_i)\Gamma(k_j)}=\frac{(k_i-1)!!(k_j-1)!!}{2(k_i+k_j)(k_i-2)!!(k_j-2)!!} &  \mbox{if  } k_i \mod 2 =0, k_j \mod 2 =0\\
       0 &\quad \mbox{otherwise.  } \\
\end{cases}
\end{multline}
Similar compact expressions cannot be found for the $J_{ij}$'s in Eq. \eqref{eq:Jij}. However, 
we see that $J_{11}=1-J_{13}, J_{12}=J^*_{14}$ and that in general we have for the $J_{ij}$'s the following:
\begin{eqnarray}
J_{21} = J^*_{12}; ~~ J_{22}=  J_{11}; ~~ J_{23} =  J_{12}; ~~ J_{24}= J_{13}\cr\cr
J_{31} = J_{13}; ~~ J_{32}=  J^*_{12}; ~~ J_{33} =  J_{11}; ~~ J_{34}= J_{12}\cr\cr
J_{41} = J_{12}; ~~ J_{42}=  J_{13}; ~~ J_{43} =  J^*_{12}; ~~ J_{44}= J_{11}.
\end{eqnarray}

\end{appendices}

\end{document}